\documentclass[11pt,a4paper]{article}
\usepackage{jheppub}

\usepackage[T1]{fontenc}
\usepackage{amsmath,amsfonts,amsbsy,amssymb,enumerate,array,accents}
\usepackage{lmodern}
\usepackage{verbatim,setspace}

\usepackage{scrpage2}

\usepackage[normalem]{ulem}

\newcommand{\Scal}[1]{\Bigl ({#1} \Bigr )}

\def\ie{{\it i.e.}\ }

\DeclareMathAlphabet{\mathpzc}{OT1}{pzc}{m}{it}
\usepackage{dsfont}

\def\eq#1{(\ref{#1})}

\def\cN{{\mathcal{N}}}
\def\cW{{\mathcal{W}}}

\newcommand{\Pp}{\Phi_{\scriptscriptstyle +}}
\newcommand{\Pm}{\Phi_{\scriptscriptstyle -}}
\newcommand{\Ppm}{\Phi_{\scriptscriptstyle \pm}}

\def\bea{\begin{eqnarray}}
\def\eea{\end{eqnarray}}
\def\be{\begin{equation}}
\def\ee{\end{equation}}

\newcommand{\CR}{\nonumber \\*}
\def\nn{\nonumber}
\def\cH{{\mathcal H}}
\def\cV{{\mathcal V}}

\newcommand{\pA}{{\text{\tiny A}}}
\newcommand{\pB}{{\text{\tiny B}}}
\newcommand{\pC}{{\text{\tiny C}}}
\newcommand{\old}{{\text{\tiny (old)}}}

\def\cE{{\mathcal{E}}}

\def\ax{\alpha}
\newcommand{\nt}{n_{{\scriptscriptstyle T}}}
\newcommand{\?}{\;\!}

\usepackage{color}

\definecolor{cardinal}{rgb}{0.6,0,0}
\definecolor{darkgreen}{rgb}{0,0.4,0}
\definecolor{darkblue}{rgb}{0, 0, 0.7}

\newcommand{\Vb}{\overline{V}}

\newcommand{\centerA}{{\large \textsc{a }}}

\makeatletter
\newcommand{\thickhline}{%
    \noalign {\ifnum 0=`}\fi \hrule height 1.3pt
    \futurelet \reserved@a \@xhline
}

\DeclareMathSymbol{\medhatsym}{\mathord}{largesymbols}{"62} 

\DeclareMathSymbol{\medtildesym}{\mathord}{largesymbols}{"65}
\newcommand\lowermedtildesym{
  \text{\smash{\raisebox{-1.2ex}{%
    $\medtildesym$}}}}
\newcommand\medtilde[1]{
  \mathchoice
    {\accentset{\displaystyle\lowermedtildesym}{#1}}
    {\accentset{\textstyle\lowermedtildesym}{#1}}
    {\accentset{\scriptstyle\lowermedtildesym}{#1}}
    {\accentset{\scriptscriptstyle\lowermedtildesym}{#1}}
}

\usepackage{datetime}

\title{Bolting Multicenter Solutions}

\preprint{IPhT-T16/150,~\,CPHT-RR048.112016}

\author[a]{Iosif Bena,}  \author[b]{Guillaume Bossard,}  \author[a]{Stefanos Katmadas}  \author[a]{and  David Turton}
\affiliation[a]{
Institut de Physique Th\'eorique, Universit\'e Paris Saclay, CEA, CNRS, \\ 
\hspace{.14cm} 91191 Gif-sur-Yvette Cedex, France}
\affiliation[b]{Centre de Physique Th\'eorique, Ecole Polytechnique, CNRS, Universit\'e Paris-Saclay, \\
91128 Palaiseau Cedex, France}

\emailAdd{iosif.bena[at]cea.fr}
\emailAdd{guillaume.bossard[at]polytechnique.edu} 
\emailAdd{stefanos.katmadas[at]cea.fr}
\emailAdd{david.turton[at]cea.fr}

\abstract{
We introduce a solvable system of equations that describes non-extremal multicenter solutions to six-dimensional ungauged supergravity coupled to tensor multiplets.
The system involves a set of functions on a three-dimensional base metric.
We obtain a family of non-extremal axisymmetric solutions that generalize the known multicenter extremal solutions, using a particular base metric that introduces a bolt.
We analyze the conditions for regularity, and in doing so we show that this family does not include solutions that contain an extremal black hole and a smooth bolt.
We determine the constraints that are necessary to obtain smooth horizonless solutions involving a bolt and an arbitrary number of Gibbons--Hawking centers. 
}

\begin{document}

\maketitle


\section{Introduction and Discussion}

The black hole information paradox~\cite{Hawking:1976ra,Mathur:2009hf} 
is a sharp and intriguing consistency challenge for any theory of quantum gravity.
String Theory offers a microscopic interpretation of black hole entropy as an enumeration of an exponentially-large number of microstates of the black hole~\cite{Strominger:1996sh}. 
It is natural to ask what the gravitational description of individual microstates is, and whether microstates have non-trivial structure on horizon scales, thus providing quantum ``hair'' for the black hole.

For extremal black holes, it has been shown that certain (coherent or semi-classical) microstates have classical descriptions that are smooth, globally hyperbolic supergravity solutions. These horizonless solutions have the same mass, charge and angular momenta as black holes with a classically-large horizon area, and are known as ``microstate geometries'', ``black hole solitons'', or ``fuzzball solutions''~\cite{Lunin:2001jy,Bena:2005va,Berglund:2005vb,Bena:2006kb,Bena:2010gg,Lunin:2012gp,Giusto:2012yz,Bena:2015bea,Bena:2016ypk}. 
For the two-charge small supersymmetric black hole, such supergravity solutions (and limits thereof) provide, upon quantization, a basis for the full space of black hole microstates~\cite{Lunin:2001jy,Lunin:2002iz,Rychkov:2005ji,Kanitscheider:2007wq}, and it has been argued that the same may be true of the three-charge large supersymmetric black hole~\cite{Bena:2014qxa}. Of course, even when there is a basis of solutions described by smooth horizonless supergravity solutions, typical microstates are complicated quantum superpositions of such basis states.

These supergravity constructions rely on the property that both for BPS \cite{Behrndt:1997ny,Bates:2003vx,Gutowski:2004yv,Bena:2004de,Gauntlett:2004qy} and for extremal non-BPS systems \cite{Goldstein:2008fq,Bena:2009en,Bossard:2011kz}, the supergravity equations of motion reduce to solvable systems of linear equations,
to which solutions can be found relatively straightforwardly.
Constructing such families of solutions for non-extremal black holes is much more complicated, as it involves solving several coupled second-order non-linear PDEs which, in the absence of supersymmetry or extremality, do not have any a priori reason to factorize. Hence, despite its importance for resolving the information paradox and investigating the experience of infalling observers~\cite{Almheiri:2012rt,Mathur:2012jk,Almheiri:2013hfa,Mathur:2013gua}, building structure at the horizon of non-extremal black holes has proven much more difficult.\footnote{There are some alternative approaches, including the construction of near-extremal microstates using probe antibranes~\cite{Bena:2011fc,Bena:2012zi}, investigating  string production near black hole horizons \cite{Dodelson:2015toa,Puhm:2016sxj}, and investigating the physics of soft particles~\cite{Hawking:2016msc}. However, such approaches lack either the generality or the precision and control that fully-backreacted supergravity solutions offer.}

The first non-extremal horizonless microstate solutions were found by Jejjala, Madden, Ross and Titchener (JMaRT) \cite{Jejjala:2005yu}, and involve a single topologically-nontrivial three-cycle, that forms a smooth bolt in the core of the solutions. These solutions have both of their angular momenta larger than those of classical black hole solutions, and 
decay via ergoregion emission~\cite{Cardoso:2005gj}. In a near-BPS limit, the solutions
have a large AdS$_3 \times S^3$ region, with the ergoregion deep inside the throat; the ergoregion emission exactly matches the Hawking radiation emitted by the holographically-dual CFT states~\cite{Chowdhury:2007jx,Chakrabarty:2015foa}.
The JMaRT solutions were found by taking certain limits of the general Cvetic--Youm family of solutions~\cite{Cvetic:1996xz}, and unfortunately this procedure does not directly enable more general constructions. Hence, for almost ten years there was little progress in this direction, except for some artisanal constructions \cite{Bena:2009fi,Bobev:2009kn,Compere:2009iy}.

The first glimmer of hope that a systematic way to build non-extremal solutions might exist appeared two years ago, when two of the present authors found a solvable system that allows a layer-by-layer construction of non-extremal supergravity solutions \cite{Bossard:2014yta, Bossard:2014ola}, allowing for multi-center generalizations of the JMaRT \cite{Jejjala:2005yu} and running-Bolt \cite{Bena:2009fi} solutions. This layered structure is a nontrivial generalization of the corresponding natural structures for supersymmetric and non-supersymmetric extremal solutions based on nilpotent subalgebras \cite{Bossard:2009my,Bossard:2009at,Bossard:2010mv,Bossard:2011kz}.

Using this graded system, the first non-extremal horizonless solution that contains two topologically-nontrivial three-cycles (or ``bubbles'') was recently constructed~\cite{Bena:2015drs}. The construction adds a Gibbons--Hawking center to the JMaRT solution, at a finite distance from the bolt, which gives rise to an additional three-cycle. This two-bubble construction also succeeded in lowering one of the two angular momenta below the black-hole bound, while the second angular momentum remained slightly over-rotating.

The system of \cite{Bossard:2014ola} therefore appears to be the tool of choice for constructing smooth horizonless solutions with non-extremal black hole charges.\footnote{
JMaRT solutions have also recently been constructed using inverse scattering methods~\cite{Katsimpouri:2014ara}, which, though currently less developed, offer another promising route to finding multicenter nonextremal solutions.}  However, this system is quite cumbersome to solve in the form in which it was originally derived.

The purpose of this paper is to de-mystify this system by rewriting all its equations in terms of new variables that simplify the differential equations, and to find a general family of axisymmetric solutions that represents a non-extremal extension to the general axisymmetric BPS and almost-BPS multicenter solutions.

The configurations described in this paper are solutions to six-dimensional $\cN=(1,0)$ supergravity coupled to $\nt$ tensor multiplets, with three commuting isometries. Upon dimensional reduction, these configurations become solutions to five-dimensional supergravity coupled to $\nt+1$ vector multiplets, with symmetric scalar manifold isometry group $SO(1,1)\times SO(1,\nt)$. Our new variables also have the advantage of making this symmetry manifest.

The new system of equations has four layers, and can be thought of as a deformation of the BPS and almost-BPS systems by additional functions that describe the deviation from extremality. 
We construct a general family of solutions in which the ansatz functions contain poles along an a priori singular three-dimensional surface and an arbitrary collection of isolated centers. 
This three-dimensional surface is similar to that appearing in the general Cvetic--Youm family of solutions~\cite{Cvetic:1996xz}, and which can be made into a smooth bolt for certain values of the parameters~\cite{Jejjala:2005yu}. 
The existence of this bolt distinguishes our solutions from the BPS and almost-BPS families, to which our solutions reduce upon taking the appropriate extremal limits.  

In the extremal systems, the poles of the ansatz functions can be chosen in such a way as to allow for finite-size regular black hole (or black ring) horizons. One can ask whether the present system contains similar solutions involving finite-size regular black objects together with a smooth bolt. As a by-product of our general regularity analysis, we show that no such solutions exist.
This is a highly nontrivial result, given that our system has the same structure as the BPS and almost-BPS systems.
It would be interesting to understand whether this is an accidental feature of the particular system of equations we use, or is rather a consequence of a deeper reason for non-existence of extremal black holes in non-extremal solutions, as we will discuss momentarily.\footnote{Note that the system constructed in \cite{Bossard:2014ola} explicitly forbids asymptotically four-dimensional non-extremal black holes, as the Noether charges of those black holes lie outside the duality orbits allowed by that system. However, a priori this does not rule out asymptotically five-dimensional solutions.}

Much like in the BPS and almost-BPS solutions, generic values of the parameters appearing in the ansatz lead to solutions with curvature singularities. Smooth horizonless solutions can be obtained by imposing certain constraints on these parameters. The resulting metric has similar behaviour near the poles of the bolt and near the added Gibbons--Hawking centers as the two-bubble solution of \cite{Bena:2015drs}, with additional parameters allowed by the more general solution of this paper.
These geometries are supported by fluxes on the bolt, on the cycles between the bolt and the Gibbons--Hawking centers, and also on the cycles between all the pairs of centers. Although these cycles are not all homologically independent, the corresponding fluxes are not additive\footnote{The flux on a cycle linking points A and B is not the sum of the fluxes on the cycles linking A and C and linking C and B with appropriate signs.} because the three-spheres that shrink are different at each Gibbons--Hawking center.

Finally, we impose absence of closed time-like curves 
near the special points of the solution, and construct the so-called ``bubble equations'' arising from these conditions. These equations have a similar, but considerably more complicated form compared to the corresponding bubble equations for extremal solutions. As in extremal solutions, these equations restrict the positions of the various centers. 

Our local analysis suggests that a large class of such solutions with arbitrary many centers should exist, although in this paper we will not explicitly solve the full set of positivity and integrality conditions to construct new explicit solutions. (An explicit example is already provided by the the two-bubble solution of \cite{Bena:2015drs}, where a complete smoothness analysis was performed.) However, the existence of a structure similar to the bubble equations for extremal solutions makes us optimistic that many more smooth multicenter solutions exist in this system, and it is interesting to anticipate what kind of physics might arise from such smooth multicenter non-extremal solutions.

Of particular physical importance is the possibility of constructing solutions that have angular momenta within the range of parameters corresponding to regular black holes, and that resemble a single-center black hole at large distances. For BPS solutions, these requirements are met by so-called ``scaling solutions'', which are solutions for which only the ratios of the distances between centers are fixed, whereas the overall scale can (classically) be tuned arbitrarily~\cite{Denef:2002ru,Bena:2006kb}. The scaling solutions develop an arbitrarily long AdS$_2$ throat (characteristic of extremal black holes) which is capped smoothly. Furthermore, since the angular momenta arise from dyonic interactions between the fluxes, in the scaling regime one has much more control over their values~\cite{Bena:2006kb,Bena:2016ypk}.

The non-extremal microstate geometries known to date do not exhibit scaling behaviour, and carry total angular momenta that violate the black hole regularity bound \cite{Jejjala:2005yu,Bena:2015drs}. In addition, these solutions have ergoregions that are significantly larger than that of the corresponding Cvetic--Youm black hole~\cite{Cvetic:1996xz}.  It is therefore natural to ask whether non-extremal scaling solutions, with lower angular momenta, exist. 
The crucial difference between non-extremal solutions and extremal solutions is that we do not expect non-extremal solutions to display arbitrary scaling, since this would produce a throat of infinite proper length, and therefore would not resemble a non-extremal black hole throat.

One therefore expects that, if non-extremal scaling solutions exist, there should be a mechanism to enforce a truncated form of scaling behaviour.
The absence of extremal black hole horizons within our general solution may be regarded as a positive indication of the existence of such a mechanism: 
if such a solution existed, one would expect to find similar horizonless solutions in which the extremal black hole horizon is replaced locally by a corresponding smooth scaling solution, with an arbitrarily long throat.

We therefore expect that solutions to the non-extremal bubble equations presented in this paper should include families that display truncated scaling behaviour between the Gibbons--Hawking centers. We believe that an exploration of this physics is of central importance for the development of the microstate geometry programme for non-extremal black holes.

This paper is organized as follows. In Section \ref{sec:gen-ansatz} we directly present our new incarnation of the system of \cite{Bossard:2014ola}, giving all the supergravity fields in terms of the functions appearing in the system. We further present our general solution describing a non-extremal bolt interacting with an arbitrary number of extremal centers, and discuss the BPS and almost-BPS limits of both the general system and the solution. In Section \ref{sec:prop} we discuss the general properties of our solution, including the restrictions required for our desired asymptotics, and the analysis of potential black hole horizons arising at the various special points of the solution. In Section \ref{sec:smooth} we discuss the conditions required for smoothness near these special points, analyzing in turn the conditions for local smooth geometry and for absence of closed time-like curves. Finally, Appendix \ref{sec:map-to-old} provides the map from the original version of the system in \cite{Bossard:2014ola} to the one described in the present paper, while Appendix \ref{sec:vec-fields} contains the explicit expressions for the vector fields arising from the general multi-center solution given in the main text.

\newpage

\section{The supergravity ansatz}
\label{sec:gen-ansatz}

In Section \ref{sec:system} we present the general structure of our system of differential equations describing solutions to six-dimensional supergravity. In Section \ref{sec:solution} we then give the general solution involving a single bolt and a set of arbitrarily many centers. We provide a short discussion of the extremal limits of the system in Section \ref{sec:extr-limits}.

\subsection{The theory and the equations}
\label{sec:system}

We consider solutions to six-dimensional $\cN=(1,0)$ supergravity coupled to $\nt$ tensor multiplets. The field content of this theory is the metric, $\nt+1$ twisted self-dual two-form potentials $C_a$, and $\nt$ scalar fields parametrized by $\nt+1$ scalars, $t_a$, subject to a quadratic constraint, where we use the non-standard numbering $a,b=1,2,4,5,\dots, \nt+2$ (the index $3$ is reserved for later convenience, since $3$ is naturally a distinguished index when $\nt=1$).

In the later parts of this paper we will focus our attention on the model containing a single tensor multiplet ($\nt=1$), whose field content reduces to a single unrestricted two-form potential, $C=C_1$, and a scalar, $\phi$, viewed as the Type IIB dilaton. Upon reduction to five and four dimensions this model gives rise to the familiar STU model. For the time being, we emphasize that we keep $\nt$ general.

Upon reduction on a circle, one obtains five-dimensional minimal supergravity
 coupled to $\nt+2$ vector multiplets, which we label by the index $I$,
using the standard numbering $I=1,2,3,\dots \nt+2$. 
We are interested in constructing smooth horizonless solutions that correspond to microstates of generic non-extremal black holes in these five-dimensional theories (or black strings in the above six-dimensional theories). 
We will focus on solutions that are asymptotically flat in five dimensions, and asymptotically $\mathds{R}^{4,1}\times S^1$ in six dimensions.

For the general six-dimensional model with $\nt$ tensor multiplets, the 
five-dimensional theory is described by totally symmetric structure 
coefficients, $C^{IJK}$, of a particular type, defined as follows. 
Let $\eta^{ab}$ be the (mostly negative) Minkowski metric of $SO(1,\nt)$,
with the following non-zero entries:
\begin{equation}
\eta_{a b} :
     \begin{cases}
        \eta_{12}=\eta_{21}=1, \\
       \eta_{ab}=-\delta_{ab} & \text{for }a,b=4,\dots \nt+2 \, . 
     \end{cases}
\end{equation}
Then the structure coefficients $C^{IJK}$ are defined by requiring that for all vectors $H_I$, we have
\begin{equation}
 \frac{1}{6}\? C^{IJK} H_I H_J H_K = \frac{1}{2}\? \eta^{ab} H_a H_b H_{3}  \label{Etaab}
\end{equation}
We also define the function $|H|$ via
\be |H|^2 = \frac12 \eta^{ab} H_a H_b \ . \ee
Note that $|H|^2$ is not strictly positive for arbitrary $H_a$, 
but the latter can be restricted such that it is. 
For the STU model considered in later sections, one simply has 
$\eta_{12}=1$, $C^{123}=1$, and all components not related by 
symmetry equal to zero.

To construct non-supersymmetric solutions to this theory, we use the partially-solvable system of differential equations of \cite{Bossard:2014ola}, whose solutions automatically solve the equations of motion of supergravity. However, the parametrization of the system appearing in \cite{Bossard:2014ola, Bena:2015drs} was rather complicated, thus making it hard to find explicit solutions in a systematic way. To remedy this, we introduce a new parametrization of the same system, resulting in a much more systematic form of the differential equations. As an added bonus, this new version of the system makes manifest the symmetries of the models based on \eqref{Etaab}, which are also present in the extremal systems of solutions to the same theory, both BPS and almost-BPS alike. Here, we concentrate on the new parametrization directly; in Appendix \ref{sec:map-to-old} we give the explicit change of variables from the version of the system presented in \cite{Bossard:2014ola}.

 
The new system of equations involves $2\nt\!+\!7$ functions on a three-dimensional base space, of which two functions, $V$, $\Vb$, can be thought as specifying an auxiliary four-dimensional Ricci-flat gravitational instanton with an isometry. Unlike in the Floating Brane ansatz~\cite{Bena:2009fi}, the full metric of this instanton does not appear in our six-dimensional metric; only the three-dimensional base metric appears as a warped component of the six-dimensional metric. An additional $\nt + 2$ pairs of functions, $K^I$, $L_I$, can be thought as parametrizing the $\nt + 2$ vector multiplets in five dimensions. The remaining function, $M$, corresponds to an angular momentum.

The three-dimensional base space metric, $\gamma_{ij}$, and the functions, $V$, $\Vb$,  are altogether a solution to the following nonlinear system of differential equations:
\begin{gather}
\Delta V = \frac{2\, \Vb}{1+ V\Vb}\,\nabla V\!\cdot\! \nabla V \,, \qquad 
\Delta \Vb ~=~ \frac{2\, V}{1+V \Vb}\,\nabla \Vb\!\cdot\! \nabla \Vb\,, \cr
R(\gamma)_{ij} = -\frac{ \partial_{(i\,}\!V \,  \partial_{j)}\!\Vb  }{( 1+V\Vb )^2} \,, 
\label{eq:R-base}
\end{gather}
describing a four-dimensional gravitational instanton. The general solution to the Euclidean Einstein equations with one isometry is of course not known, but starting from any known instanton solving  \eqref{eq:R-base}, one obtains a solvable system of equations in this auxiliary base space. In particular, the Laplacian, $\Delta$, appearing in the remainder of this section is the one computed using the metric $\gamma_{ij}$.

The equations for the rest of the functions that determine the solution then become
\begin{eqnarray}\label{eq:Lapl-eqns-gen}
\Delta K_I &=& \frac{2\, V }{1+V \Vb}\,\nabla \Vb \!\cdot\! \nabla K_I\,, \cr
\Delta L^I &=& \frac12\,\frac{V}{1+V \Vb}\,C^{IJK}\,\nabla  K_{J} \!\!\; \cdot\! \nabla K_{K} \,,  \\
\Delta M  &=&  \nabla \!\cdot\!\left( \frac{V}{1+V\Vb}\, \left( L_I \nabla K^I -2\, M \nabla \Vb \right) \right) \,,
\nonumber
\end{eqnarray}
where the structure constants $C^{IJK}$ are given in \eqref{Etaab}. When solved in the order outlined above these equations are linear, and therefore represent a solvable system on the base specified by a solution to \eqref{eq:R-base}. 

Any solution to the system \eqref{eq:Lapl-eqns-gen} gives rise to a metric, two-forms and scalar fields that solve the supergravity equations of motion. 
The six-dimensional Einstein-frame metric is given in terms of a function, $W$, a vector of functions, $H_I$, and three vector fields, $A^3$, $k$ and $w^0$. Anticipating our focus on asymptotically-flat solutions in five dimensions in the next section, we write the metric as:
\begin{equation}\label{eq:6D-metr}
ds^2 = \frac{H_3}{|H|} ( dy + A^3)^2 - \frac{W}{H_3|H|} ( dt + k )^2 
       +|H| \Scal{ \frac{1}{W} ( d\psi + w^0 )^2 + \gamma_{ij} dx^i dx^j} \,.
\end{equation}
The notation for the Kaluza--Klein vector field $A^3$ is motivated by the fact that it is one of the gauge fields appearing symmetrically in the STU model in the five-dimensional theory. The vectors $A^3$ and $k$ decompose as
\begin{equation}\label{eq:6d-KK}
A^3 ~=~ A^3_t\, (dt + \omega) + \ax^3\,(d\psi + w^0) +  w^3 \,, \qquad\quad k ~=~ \frac{\mu}{W}\,(d\psi + w^0)+\omega  \,,
\end{equation}
where $A^3_t$, $\ax^3$, $\mu$ and $w^0$, $\omega$ are three scalars and two vector fields on the three-dimensional base.
The functions $W$, $\mu$, $H_I$ appearing in the metric are given in terms of the functions  $(V,\bar{V},K_I,L^I,M)$ as follows:
\begin{align}\label{eq:scal-facts}
 W = &\, \left(  (1 + \Vb)\, M - \frac12 \, K_I L^I +  \frac{1}{24}\,\frac{V}{1+V\Vb}\, C^{IJK}\,K_I K_J K_K \right)^2  \nonumber\\
&+\frac{1-V}{1+V\Vb}\,\left( \frac16\, C^{IJK} K_I K_J K_K\, M 
+ \frac13\, (1 + \Vb)  C_{IJK}\,L^I L^J L^K 
 - \frac14\, C^{IJK}K_J K_K\, C_{ILM}L^L L^M \right) \,,
 \nonumber\\
H_I = &\, \frac12\,C_{IJK}\,L^J L^K - K_I\, M + \frac12\,\frac{V}{1+V\Vb}\,\left( (K_J L^J )\,K_I - \frac{1}{2} C_{IJK}L^J\, C^{KLP}\,K_L K_P \right)  \,,
 \nonumber\\
\mu = &\, (1 + \Vb)\,M^2 -\frac12\, M\, K_I L^I 
- \left( 1 + 2\,\frac{V-1}{1+V\Vb} \right)\,C_{IJK}\,L^I L^J L^K
\CR
    &\, + \frac12\,\frac{V}{1+V\Vb}\,\left( -\frac1{12}\, C^{IJK}\,K_I K_J K_K\, M + \frac14\, C^{IJK}K_J K_K \, C_{ILM}L^L L^M \right) \,.
\end{align}
Similarly, the vector fields $\omega$, $w^0$ and $w^3$ are determined by the first-order equations
\begin{align} \label{wIEq-1} 
\star d \omega =&\, d M - \frac{V}{1+V\Vb}\, \left( L^I \,d K_I -2\, M \, d \Vb \right) ,
\CR
\star d w^0 =&\, -(1+\Vb)\, d M -\frac{1}{2}\,\frac{1-V\Vb-2\,V}{1+V\Vb}\, \left( L^I \,d K_I -2\, M \, d \Vb \right) 
+ \frac{1}{2}\,K_I \,d L^I
\CR  
&\,   - \frac1{24}\,\frac{V}{1+V\Vb}\,d\left( C^{IJK}\,K_I K_J K_K \right)
      + \frac1{24}\,\frac{C^{IJK}\,K_I K_J K_K}{(1+V\Vb)^2}\left( V^2 d\Vb + dV \right) , 
\CR
\star d w^I =&\, d L^I - \frac{1}{4}\, \frac{V}{1+V\Vb}\,d\left( C^{IJK}\,K_J K_K \right)
  + \frac{1}{4\,(1+V\Vb)^2}\,C^{IJK}\,K_J K_K \left( V^2 d\Vb + dV \right)  ,
 \end{align}
 where the Hodge star in taken in the metric $\gamma_{ij}$ and we have given the $w^I$ in an $SO(1,1)\times SO(1,\nt)$ covariant form; the $w^a$ will appear in the matter sector, as we will discuss shortly.

The $\nt+1$ scalar fields, $t_a$, are given by the expression
\begin{equation}
t_a = \frac{H_a}{|H|}  \,,
\end{equation}
with the $H_I$ as in \eqref{eq:scal-facts}. This set of constrained scalars can be decomposed into the $\nt$ physical scalar fields, namely the dilaton, $\phi$, and the $\nt-1$ real axions, $\varsigma_a$, for $a=4$ to  $\nt+2$, as
\begin{equation}
%
\qquad
t _{a} :
     \begin{cases}
        t_1 = e^\phi \,, \\
				t_2 = e^{-\phi} + \frac{1}{2} e^\phi \sum_{b}  \varsigma_{\scalebox{0.7}{$b$}}^{\, 2} \,, \\
       t_{a} = e^\phi \varsigma_a & \text{for }a =4,\dots \nt+2 \,
     \end{cases}
\end{equation}
leading to the expressions
\be e^{\phi} = \frac{H_1}{|H|} \ , \qquad \varsigma_a = \frac{H_a}{H_1} \ . \ee
The $SO(\nt)\backslash SO(1,\nt)$ coset representative is parametrized in terms of the physical scalars as 
\be V = \left(\begin{array}{ccc} e^\phi & \frac{e^\phi}{2}(   \varsigma^T\varsigma) & e^\phi \varsigma^T \\ 0& e^{-\phi} & 0\\
0&  \varsigma & \mathds{1} \end{array}\right) \ , 
\ee
where $\mathds{1}$ is the $(\nt-1)$-dimensional identity matrix, so that $V$ is a square $(\nt+1)$-dimensional matrix.
The matrix $V$ defines the symmetric $SO(1,\nt)$ matrix $M = V^T V$, which is given by 
\be M_{ab} =  \frac{H_a H_b}{|H|^2} - \eta_{ab} \,. \label{eq:6D-dil} \ee
The inverse of $M$ is $M^{ab} = \eta^{ac} \eta^{bd} M_{cd}$. 

The $\nt+1$ two-form potentials, $C_a$, give rise to one anti-self-dual and $\nt$ self-dual three-form field strengths $G_a=dC_a$, satisfying the twisted self-duality equations
\begin{equation} \label{eq:6D-eom}
M^{ab} \star_6 G_b+ \eta^{ab} G_b = 0  \, .
\end{equation}
%
The two-form potentials, $C_a$, can be expressed in terms of three-dimensional quantities. We first introduce the scalars $A_t^a$, $\beta_a$ and $\ax^a$ with the latter identified as axions in the reduction to four dimensions. Additionally, we introduce the three-dimensional one-forms $w^a$, $v_a$ and $b_a$; the $w^a$ are determined by \eqref{wIEq-1}, while $v_a$ and $b_a$ will be defined shortly. Finally, we define the two-forms in three dimensions, $\Omega_a$, through
\begin{equation} \label{eq:Om-def}
d \Omega_a = v_a \wedge d w^0 - \eta_{ab} w^b \wedge d w^3 + b_a \wedge d \omega \,.
\end{equation}
In terms of these quantities, we have
\begin{eqnarray}
C_a \,&=&\, \eta_{ab}\? A_t^b\, (dy + w^3 ) \wedge(dt+\omega) + \eta_{ab}\? \ax^b\, (dy + w^3)  \wedge (d\psi + w^0)
-\beta_a\,(dt+\omega)  \wedge (d\psi + w^0) \quad \CR
&&{} - \eta_{ab}\? w^b \wedge (dy + w^3 )+ b_a \wedge(dt+\omega) + v_a \wedge (d\psi + w^0 ) + \Omega_a \,.
\label{eq:2-form-exp}
\end{eqnarray}
Note that the $\Omega_a$ ensure that the field strengths, $G_a$, depend on the vectors $w^a$, $b_a$ and $v_a$ only through the gauge-invariant quantities $dw^a$, $db_a$ and $dv_a$.
The $\Omega_a$ vanish for axisymmetric solutions, since all vector fields have components only along the angular coordinate around the axis, implying that their wedge products appearing in \eqref{eq:Om-def} vanish identically. We only construct axisymmetric solutions in the current work, so we henceforth set $\Omega_a$ to zero.

The one-forms, $v_a$, $b_a$ in \eqref{eq:2-form-exp} are determined in terms of the functions appearing in the ansatz  by solving the first-order equations
\begin{align}
\label{baEq-2} 
\star db_I ~=~&\, \frac{1-V}{1+V\Vb}\,d K_I + \frac{K_I}{(1+V\Vb)^2}\,\left((V-1)\, V\, d \Vb + (1 + \Vb) d V \right)\,,
\\
\label{eq:alm-NE-mag}
 \star d v_I ~=~ &\, -\frac{V}{1+V\Vb}\,d K_I + \frac{K_I}{(1+V\Vb)^2}\left( V^2 d\Vb + dV \right)
\,,
\end{align}
where we again give a fully covariant form for completeness, even though $db_3,$ $dv_3$ are not relevant for our solution. The explicit form for these one-forms can be obtained straightforwardly for any given solution to the system (\ref{eq:R-base},\ref{eq:Lapl-eqns-gen}). The scalars $\beta_a$ are given by 
\begin{align}
\beta_a = \frac{H_a}{|H|^2}\left( L^3  - \frac12\,\frac{V}{1 + V\,\Vb}\? \eta^{bc} K_b K_c \right) \, .
\end{align}
Similarly to the vectors $w^a$, the electric potentials $A_t^a$ and axions $\ax^a$ in \eqref{eq:2-form-exp} are also extended by the scalars $A_t^3$, $\ax^3$ of \eqref{eq:6d-KK} in the five-dimensional reduction of the theory. For the STU model ($\nt=1$), on which we shall concentrate in later sections, these scalar fields for $I=1,2,3$ are given by (note that the Einstein summation convention does not apply in the following two equations)
\begin{align}
 A^I_t = & \,\frac1{2\,H_I}\left( 2\, (1 + \Vb)\, M - \sum\limits_{J} K_J L^J 
  +\frac{1}{2} \frac{V\,K_1 K_2 K_3}{1+V\Vb} - 2\,K_I L^I \frac{V - 1}{1 + V\Vb} \right)\,,
\label{eq:5dzeta}
\\
\ax^I = &\, \frac{1}{H_I}\left( M - \frac{V\, K_I L^I}{1+V\Vb}\right) .
\label{eq:5dax}
\end{align}
The corresponding expressions for these fields in more general models are straightforward to obtain.\footnote{Defining $\det H= \tfrac{1}{6}C^{IJK} H_I H_J H_K$, one must make the replacements 
\begin{align}
 \frac{1}{H_I} \rightarrow&\, \frac{1}{2 \det H}C^{IJK}H_J H_K \,,
\CR
\frac{K_I L^I}{H_I}\rightarrow&\, \frac{1}{2\det H}\? \Scal{  C^{IJK} H_J H_K\? K_L L^L + L^I \? C^{JKL}K_J H_K H_L - C^{IJL}K_J\? C_{LPQ} L^P\? C^{QRS}H_R H_S}\ .
 \end{align}
}

We close this general discussion of the system by pointing out a symmetry that was not evident in the variables used in \cite{Bossard:2014ola, Bena:2015drs}, but becomes clear in the covariant version of the system described above. For some constants, $k^I$, one can verify that the equations \eqref{eq:Lapl-eqns-gen} transform linearly among themselves under the transformation defined by
\begin{align}
K_I &\to K_I+ k_I \Vb\,, \CR 
L^I &\to L^I + \frac12\,C^{IJK}\,k_J K_K + \frac14\,C^{IJK}\,k_J k_K \Vb\,, \CR 
M &\to M + \frac12\,k_I L^I + \frac18\,\frac{V}{1+V\Vb}\,C^{IJK}\,k_I K_J K_K \CR 
&\quad
+ \frac14\,\left( 1- \frac12\,\frac{1}{1+V\Vb}\right)\, \left( C^{IJK}\,k_I k_J K_K + \frac13\,C^{IJK}\,k_I k_J k_K \Vb \right)\,.
\label{eq:gaugespectral}
 \end{align}
It then follows that one may act with this symmetry on any solution for $K_I$, $L^I$ and $M$ to obtain a new solution. It will prove useful in packaging our general solution in the next section to make use of the following invariant combinations
\begin{equation}\label{eq:L-M-invar}
L^I - \frac1{4\,\Vb}\,C^{IJK}\,K_J K_K\,, \qquad\quad
M -\frac1{2\,\Vb}\, K_I L^I + \frac1{4\,\Vb^2}\,\frac{2+V\Vb}{1+V\Vb}\,K_1 K_2 K_3\,.
\end{equation}
When acting on the vector fields this symmetry leaves the combination $d\omega + dw^0$ invariant, and transforms:
\begin{align}
dv_I &\to dv_I + k_I\, d\rho \,, \CR 
dv_I -db_I &\to dv_I -db_I - k_I\, d\sigma \,, \CR 
dw^I &\to dw^I - \frac12\,C^{IJK}\,k_J (dv_K -db_K) + \frac14\,C^{IJK}\,k_J k_K\, d\sigma\,, \CR 
dw^0 &\to dw^0 + \frac12\,k_I dw^I - \frac18\,C^{IJK}\,k_I k_J (dv_K -db_K) - \frac14\,k_1 k_2 k_2\, d\sigma  \,,
\label{eq:gauge-on-vecs}
 \end{align}
where we have used the conserved currents
\begin{equation}\label{eq:rhp-sig}
\star d\rho = \frac{\Vb\,dV - V\,d\Vb}{(1+V\Vb)^2} \,, \qquad \star d\sigma= \frac{d\Vb + \Vb^2\,dV}{(1+V\Vb)^2}\,.
\end{equation}
This symmetry is conjugate in $SO(4,3+\nt)$ to the gauge transformations/spectral flows appearing in the BPS and almost-BPS systems~\cite{Bena:2005ni,Bena:2008wt,Dall'Agata:2010dy}, via an S-duality and a change of time coordinate $t\to t - \psi$.

\subsection{The solution}
\label{sec:solution}

In this paper, we focus on solutions containing a single bolt and an arbitrary number of centers, by which we mean locations in which the ansatz functions have poles, and which can potentially become smooth Gibbons--Hawking centers for certain choices of parameters. A necessary starting point is obtaining an appropriate solution to \eqref{eq:R-base}. Throughout this paper, we work with a solution to these equations specified by choosing the three-dimensional base to be the base space of the Euclidean Kerr solution:
\bea\label{eq:3D-base}
\gamma_{ij} dx^i dx^j &=& 
\left( 1+ \frac{a^2 \sin^2 \theta}{r^2-c^2} \right) dr^2 
+ (r^2 - c^2+a^2\sin^2\theta)d\theta^2 +(r^2 - c^2)\sin^2\theta d\varphi^2 \,,
\eea
where $a$ and $c$ are real constants, and we take $a>c>0$ by convention. This is a natural choice for axisymmetric solutions above the extremality bound, but is not unique in general. 

At the locus $r=c$, the base metric $\gamma_{ij}$ is singular. In our full six-dimensional solution, this singularity can be resolved into a bolt, with two nuts at the North pole and South pole of the bolt, defined by $\cos \theta=\pm 1$ respectively (we follow the terminology of~\cite{Gibbons:1979xm}).
Such a smooth bolt is present in the JMaRT solution~\cite{Jejjala:2005yu} and the two-bubble solution of \cite{Bena:2015drs}.
%
The solutions that we consider can be thought of as adding an arbitrary number of centers to this bolt.

In order to look for explicit solutions, we restrict attention to axisymmetric solutions built on the base \eqref{eq:3D-base}. 
This implies that all centers outside the bolt at $r = c$ must lie on the symmetry axis, \ie at $\cos \theta=\pm 1$ and $r>c$.
Similarly, all vectors on the 3D base are constrained to have a single component along $\varphi$, for example
\bea
\omega = \omega_\varphi d\varphi \,, \qquad
w^I = w^I_\varphi d\varphi \,.
\eea

We now proceed to construct an explicit solution, starting with the functions $V$, $\Vb$, which are determined by \eqref{eq:R-base} once the base metric in \eqref{eq:3D-base} is used. Explicit expressions for these functions can be recovered from the Kerr solution:
\bea\label{eq:V-Vb-def}
V &=& 1 +\frac{m_-}{r-a\cos\theta} \,,  \cr
\Vb &=& \frac{a^2-c^2}{m_-}\frac{1}{ \Sigma_+} ~\equiv~ \frac{a^2-c^2}{m_- (r+a\cos\theta) +c^2-a^2},
\eea
where $m_-$ is a (real-valued) constant of integration and we defined the combination $\Sigma_+$ for later convenience. With this choice for the base metric and the functions $V$ and $\Vb$, one may proceed to solve the remaining equations in \eqref{eq:Lapl-eqns-gen} in the order in which they appear, since they become linear equations with sources involving the functions obtained by the previous steps. 

The first of \eqref{eq:Lapl-eqns-gen} is homogeneous in the $K_I$ and allows for zero modes with simple poles anywhere on the axis. Using \centerA to label a point at position $R_\pA$ along the axis, we denote by $\Sigma_\pA$ the Euclidean distance 
\begin{equation}
\Sigma_\pA = \sqrt{(r^2-c^2)\,\sin^2 \theta+(R_\pA-r\,\cos\theta)^2}\,.
\end{equation}
Then we find the solution for the $K_I$
\begin{eqnarray}\label{eq:KI-def}
K_I &\equiv& h_I + \medtilde{K}_I ~=~ 
h_I + k_I \Vb + \sum\limits_\pA \frac{2\,n_{I}^\pA}{\Sigma_+\,\Sigma_\pA} 
\left( r+a\,\cos\theta+\frac{a^2-c^2}{R_\pA-a}\,\cos\theta  \right)\,,
\end{eqnarray}
where $h_I$, $k_I$ and $n_{I}^\pA$ are integration constants and $\Sigma_+$ was defined in \eqref{eq:V-Vb-def} above. Note that the second term in $K_I$ can be introduced via the symmetry transformation \eq{eq:gaugespectral}, so that one can solve the equations without it, then re-introduce it by hand. A similar structure is present in $L^I$ and $M$; in order to parametrize this in what follows, we have introduced above the function  $\medtilde{K}_I $ which asymptotes to zero.

It turns out that a combination of the shift parameters, $k^I$, and the asymptotic constants, $h^I$ is relevant for describing the solution. We therefore introduce the shorthand notation
\begin{equation}\label{eq:qI-def}
 q_I = k_I - h_I\,,
\end{equation}
which will be used in the functions below. The parameters $q_I$ will also be convenient quantities to use in the discussion of regularity in the next section.

With this notation, the solution for the $L_I$ takes the form
\begin{eqnarray}\label{eq:LI-def}
L^I &=& \frac{C^{IJK} \medtilde{K}_{J}\medtilde{K}_{K}}{4\,\Vb} +l^I + \frac{p^{I}_-}{r+c \cos\theta} + \frac{p^{I}_+}{r-c \cos\theta} + \sum\limits_\pA \frac{P^I_{\pA}}{\Sigma_\pA}\\
&&\hspace{-2mm}- \frac{m_-}{a^2 - c^2}\,\sum\limits_{\pA,\pB} \frac{C^{IJK} n_{J}^\pA n_{K}^\pB}{\Sigma_\pA \Sigma_\pB} \,
\left( (r + a\, \cos{\theta}) - \frac{(a^2 - c^2)}{(R_\pA - a) (R_\pB - a)}\,(r -(R_\pA + R_\pB-a)\,\cos{\theta})  \right) .
\nonumber
\end{eqnarray}
Note that the first term in $L^I$ includes all terms that depend on $k_I$; this follows from the invariance of the combination \eqref{eq:L-M-invar} and can be seen to reproduce the dependence in \eq{eq:gaugespectral}. The constants $p^I_\pm$ and $P^I_\pA$ parametrize harmonic components of this solution sourced at the poles of the bolt and at the Gibbons--Hawking points respectively. Here, we chose to disregard any higher multipole harmonic functions sourced at these locations, which can in principle be added to  \eqref{eq:LI-def}. We make this restriction using intuition from the extremal multi-center solutions, BPS and almost-BPS, in which such higher order harmonic pieces in the $L^I$ are not physically relevant. 

The same comments apply to the $k_I$-dependent terms in $M$, for which we find the solution:
\begin{eqnarray}\label{eq:M-def}
M&=& \frac{\medtilde{K}_{I} L^I}{2\,\Vb} - \frac{1}{12\,\Vb^2}\,\frac{2+ V\,\Vb}{1+ V\,\Vb}C^{IJK} \medtilde{K}_{I}\medtilde{K}_{J}\medtilde{K}_{K}
+ \frac{1}{1+V\Vb}\,\left( l^0 - \frac{m_-}{2\,(a^2 - c^2)}\,\Sigma_+\,l^I \medtilde{K}_{I}\right)
\cr
&&{} +\frac{\Sigma_+}{r^2-c^2+a^2 \sin^2\theta}\,\left[ q_0 + J_+ \,\left(
\frac{2 \cos\theta}{r-c \cos\theta}-\frac{(a+c)\sin^2\theta}{(r-c \cos\theta)^2}\right) 
 \right. \cr
%
&&{}
\qquad\qquad  +J_- \,\left(
\frac{2 \cos\theta}{r+c \cos\theta}-\frac{(a-c)\sin^2\theta}{(r+c \cos\theta)^2}\right)
+\, \sum\limits_\pA q^0_{\pA}\,\frac{\left(r\, (R_\pA+ a)-\cos\theta\, \left(a\, R_\pA + c^2\right)\right)}{\Sigma_\pA} 
\cr
&&{}\left. \qquad \qquad 
+\sum\limits_\pA J^\pA  \frac{(r^2 - c^2) (\cos\theta\,R_\pA- r) + a \,\sin^2\theta (r\,R_\pA - c^2 \cos\theta)}{\Sigma_\pA^3} 
\right]\cr
&&{} 
%
%
+\sum\limits_{\pA,\epsilon=\pm} \frac{p^{I}_{\epsilon} n_{I}^\pA}{(R_\pA - a)\,\Sigma_\pA} 
\left[ \frac{m_-}{a+{\epsilon}\;\! c} - \frac{m_-\, \Sigma_+\,(R_\pA-a)}{(a^2-c^2)\,(r-{\epsilon}\;\! c \cos\theta)}
+ \left( \frac{a+{\epsilon}\;\! c}{a-{\epsilon}\;\! c} \right) \frac{a-{\epsilon}\;\! c - m_-\cos\theta}{r-{\epsilon}\;\! c \cos\theta} \right.\cr
&&{} 
\left.\qquad\qquad\qquad\qquad
-\frac{2}{a -{\epsilon}\;\!  c}\,\frac{a\, (a -{\epsilon}\;\!  c)\, (r+{\epsilon}\;\! c\, \cos\theta) - a\, m_- (r\,\cos\theta+{\epsilon}\;\! c) + m_- a^2 \sin^2\theta}{r^2-c^2+a^2 \sin^2\theta}\, \right]
\cr
%
%
&&{} 
+\sum\limits_{\pA,\pB} \frac{n_{I}^\pA P^I_{\pB} }{R_\pA-a} \,\frac1{\Vb\,\Sigma_\pA\,\Sigma_\pB}\times
\cr
&&{} \qquad\qquad
\left[ -(R_\pA-a) -\frac{a^2-c^2}{r^2-c^2+a^2 \sin^2\theta}\,
\left( a^2 \sin^2\theta + \frac{(\Sigma_\pA-\Sigma_\pB)^2-(R_\pA-R_\pB)^2}{2\,(R_\pA-R_\pB)} \right) \right]\cr
&&{} 
+\frac{m_-}{3 \Vb\,(1+V\,\Vb)}\,\sum\limits_{\pA,\pB,\pC} \frac{C^{IJK}n_{I}^\pA n_{J}^\pB n_{K}^\pC}{\Sigma_\pA\,\Sigma_\pB\,\Sigma_\pC} \times
\cr
&&{} 
\left\{ (2+\Vb+V\,\Vb)\,\left[\frac{r+a\, \cos\theta}{a^2-c^2}
+ \cos\theta \left( \frac{1}{R_\pA-a} + \frac{1}{R_\pB-a} + \frac{1}{R_\pC-a} \right) \right] \right.\cr
&&{} 
-\frac{(R_\pA + R_\pB + R_\pC - 3 a) (r-a \;\! \cos\theta ) + (a^2 - c^2) \cos\theta}{(R_\pA-a)(R_\pB-a)(R_\pC-a)}
\left(1 + \cos^2\theta\,\Vb + V\,\Vb\,\sin^2\theta \right) \cr
&&{} 
\left. + \frac{2\,m_-a}{(R_\pA-a)\,(R_\pB-a)\,(R_\pC-a)}\,\Vb\,\cos2\theta \right\} .
\end{eqnarray}
In the above, the term that contains $(R_\pA-R_\pB)$ in the denominator should be understood to be zero when {\large \textsc{a}}$=${\large \textsc{b}}. Here, the constants $l^0$, $q_0$, $J_\pm$, $J^\pA$ and $q^0_{\pA}$ parametrize zero modes for $M$.

We close this section by forewarning the reader that we impose a redefinition of the $P^I_\pA$ in the following sections and in Appendix \ref{sec:vec-fields}, in order to simplify expressions. Explicitly, we set
\begin{equation}\label{eq:p-redef}
P^I_{\pA} = C^{IJK}\, n_{J}^\pA (p_K^{\pA}-q_K) \,,
\end{equation}
where the $p_I^{\pA}$ are triplets of constants at each Gibbons--Hawking center. While this does not impose any restriction for general $n_I^{\pA}$, $p_I^{\pA}$, the redefinition \eqref{eq:p-redef} is particularly useful when considering vectors $n_I^{\pA}$ of restricted rank, as we shall see later.

\subsection{Extremal limits}
\label{sec:extr-limits}

In view of the manifest $SO(1,1)\times SO(1,\nt)$ symmetry, the present system lends itself easily to comparison with the BPS and almost-BPS systems. In order to obtain an extremal limit, one must ensure that the three-dimensional base of the metric is flat, which implies that the Ricci tensor in \eqref{eq:R-base} must vanish. There are two ways of obtaining this result, namely setting either $V$ or $\Vb$ to a constant.

In the explicit solution in Section \ref{sec:solution}, $\?\Vb$ can be made constant while keeping $V$ non-trivial only by holding $m_-$ fixed and non-zero and taking the limit $a \to c$, in which case $\?\Vb$ becomes zero.
Alternatively, $V$ can be made constant while keeping $\?\Vb$ non-trivial only by sending $m_-$ to zero and $a\to c$, keeping the ratio $(a^2-c^2)/m_-=p^0$ fixed. In this case $V$ becomes equal to 1.
In both extremal limits, since $a \to c$, the metric \eqref{eq:3D-base} degenerates. 

Upon setting $\Vb$ to a constant, one finds that the defining equations \eqref{eq:Lapl-eqns-gen} reduce to the
almost-BPS system as given in \cite{Goldstein:2008fq}, upon identifying the combination $V/(1+V\,\Vb)$ as a
harmonic function. In the explicit solution in Eq.\;\eq{eq:V-Vb-def}, we have $\Vb=0$, and $V$ is harmonic with a single pole at $r-c\,\cos\theta$. The $K^I$ become harmonic, as can be seen directly from Eq.\;\eq{eq:Lapl-eqns-gen}, or by setting $a=c$ in \eqref{eq:V-Vb-def} and \eqref{eq:KI-def}. The remaining functions, $L_I$ and $M$, as
given by \eqref{eq:LI-def}--\eqref{eq:M-def}, are consistent with the solution to the almost-BPS equations with
a single pole in $V$, as given in \cite{Bena:2009ev,Bena:2009en}. However, the embedding of the various functions in the
supergravity solution described by \eqref{eq:scal-facts} is not the standard one; rather, it is related to the one
in \cite{Goldstein:2008fq,Bena:2009ev,Bena:2009en} by a four-dimensional S-duality and a gauge transformation.

Similarly, setting $V$ to a constant simplifies in a different way the defining equations \eqref{eq:Lapl-eqns-gen}, this time
leading to the BPS system. Setting  $V=1$ for definiteness, and introducing the notation $\cH^\Lambda$, $\cH_\Lambda$ for $\Lambda=0,I$ for the BPS functions that form a symplectic vector of functions, 
one finds the following change of variables:
\begin{gather} 
 \Vb = \frac{2}{\cH_0} - 1\,, \qquad K_I = -2\,\frac{\cH_I}{\cH_0}\,, \qquad 
 L^I = \cH^I +\frac{1}{2\,\cH_0}C^{IJK} \cH_J\cH_K\,,
 \CR
 M = -\frac{1}{2}\left(\cH_0 \cH^0 + \cH_I \cH^I \right)- \frac{1}{\cH_0}  \cH_1 \cH_2\cH_3\,.
 \label{eq:BPS-har}
\end{gather}
In terms of the explicit solution in Section \ref{sec:solution}, in the BPS limit $\cH_0$ has a single pole at $r+c\,\cos\theta$, while the remaining harmonic functions are those of a standard BPS smooth solution. 
In this limit, defining $dv_0$ to be the BPS limit of $-2\, d\sigma$ in Eq.\;\eqref{eq:rhp-sig}, we have
\begin{equation}
\begin{pmatrix} \star dw^\Lambda \\ \star dv_\Lambda \end{pmatrix} = \begin{pmatrix}  d\cH^\Lambda \\ d\cH_\Lambda \end{pmatrix}\,, 
\end{equation}
and 
\begin{equation}
\star d\omega = 
\frac{1}{2}\,\left( \cH^\Lambda d\cH_\Lambda - \cH_\Lambda d\cH^\Lambda \right)\, . 
\end{equation}
With these definitions, the symmetry in \eqref{eq:gaugespectral} survives and its action on the vector fields and harmonic functions is conjugate to a spectral flow transformation with parameters $-\tfrac12\,k_I$, through a gauge transformation in five dimension that amounts to the redefinition $\cH_0\to\cH_0 -2$. We observe that this is consistent with the transformation \eqref{eq:rhp-sig}, noting that $b_I$ vanish identically in the BPS limit.

\section{General properties of the solution}
\label{sec:prop}

In this section we analyze the local regularity conditions on the parameters of the general solution in Section \ref{sec:solution}, focusing on the various interesting locations, namely asymptotic infinity, the centers away from the bolt, and the bolt itself. 
As mentioned above, we are interested in microstates of black holes in five dimensions (and black strings in six dimensions) and so we are interested in solutions with $\mathds{R}^{4,1}\times S^1$ asymptotics.

We first consider the behaviour of the solution near asymptotic infinity in Section \ref{sec:infinity}, identifying the appropriate constraints. We then proceed in Section \ref{sec:no-bh} to analyze the possibility of obtaining regular black hole horizons at any of the special points in the bulk, namely the poles of the bolt and the centers away from the bolt, and show that such regular horizons cannot be built using our ansatz, unless one takes an extremal limit. 

\subsection{Asymptotics}
\label{sec:infinity}

In the solution that is obtained by directly substituting \eqref{eq:KI-def}--\eqref{eq:M-def} in the relevant expressions, various components of the metric and fields tend to non-zero constants at asymptotic infinity. 
In order to obtain standard asymptotics, we first make a set of gauge transformations and coordinate transformations to set these constants to zero.
These operations do not impose any constraints on the parameters of the general solution.


We start by shifting away the asymptotic constants from the off-diagonal components of the metric and the two-forms $C_a$, using a set of diffeomorphisms and gauge transformations. Specifically, one can shift to zero the asymptotic values of the scalars $\ax^a$, $\beta_a$ and $A_t^a$ in \eqref{eq:2-form-exp} by a gauge transformation on the two-forms, provided that one redefines the vector fields as
\begin{align}\label{eq:redef1}
 w^a{}' = &\; w^a + A_t^a \big|_{\scriptscriptstyle\infty}\omega + \ax^a \big|_{\scriptscriptstyle\infty} w^0\ , 
\CR
 v_a' = &\; v_a - \beta_a \big|_{\scriptscriptstyle\infty} \omega 
                         + \eta_{ab}\,\ax^b \big|_{\scriptscriptstyle\infty} w^3 \ , 
\CR
 b_a' = &\; b_a + \eta_{ab}\,A_t^b \big|_{\scriptscriptstyle\infty} w^3
                         + \beta_a \big|_{\scriptscriptstyle\infty} w^0\ , 
\end{align}
where primes denote redefined quantities, we denote asymptotic values by $\big|_{\scriptscriptstyle\infty}$,  and we use \eqref{Etaab}. 
Having done these redefinitions, we immediately drop the primes on the above expressions, and likewise for the following two steps.

Next, one may remove the asymptotic constants of $A_t^3$ and $\ax^3$ that appear in the Kaluza--Klein gauge field $A^3$ given in \eqref{eq:6d-KK}, by a diffeomorphism that mixes the coordinate $y$ with $t$ and $\psi$ at infinity, provided that one makes the redefinitions
\begin{align}\label{eq:redef2}
 v_a' = &\; v_a + \ax^3 \big|_{\scriptscriptstyle\infty}\eta_{ab}\, w^b \ , 
\CR
 b_a' = &\; b_a + A_t^3 \big|_{\scriptscriptstyle\infty} \eta_{ab}\,w^b  \ , 
\CR
 \beta_a'  = &\; \beta_a +\ax^3\big|_{\scriptscriptstyle\infty} \eta_{ab}\, A_t^b\ . 
\end{align}
Additionally, one can shift away the constant values of $\omega$, $w^3$ and the $w^a$ at infinity by making an appropriate diffeomorphism that mixes the coordinates $t$, $y$ with $\varphi$, as well as by doing a further gauge transformation on the two-forms; these do not induce any additional redefinitions. A final redefinition we use is a diffeomorphism mixing time with one of the compact directions, $t = t' + \gamma\, \psi$, and introducing the redefined fields
\begin{eqnarray}\label{eq:t-psi-mix}
\omega' = \omega - \gamma\, w^0 \,, \qquad 
\mu' = \mu + \gamma\, W \,,\qquad 
\alpha^I{}' = \alpha^I + \gamma\, A_t^I\,,\qquad 
v_I' = v_I + \gamma\, b_I \,
\end{eqnarray}
where the value of $\gamma$ will be determined by the asymptotic conditions below.
We again immediately drop the primes on all the above expressions.

The concrete expressions for the various asymptotic constants appearing in the above redefinitions are straightforward to obtain using the solution given in Section \ref{sec:solution}, but are not illuminating and play no role in the following. Therefore, we refrain from giving them explicitly and henceforth work with the quantities after \eqref{eq:redef1} and \eqref{eq:redef2} have been applied.

We next discuss the conditions arising from the asymptotics that impose constraints on the parameter space. For simplicity we shall consider only one tensor multiplet; the generalization to arbitrary $\nt$ is straightforward, but requires the introduction of a unit norm vector of $SO(1,\nt)$. To obtain our desired $\mathds{R}^{4,1}\times S^1$ asymptotics, we impose the fall-off behaviour
\bea\label{eq:as-beh}
W = \frac{1}{r^2}+ {\cal O}\Scal{\frac{1}{r^3}}\,, \qquad 
H_I = \frac{1}{r} +{\cal O}\Scal{ \frac{1}{r^2}}\,, \qquad 
\mu= {\cal O}\Scal{\frac{1}{r^3}}\,.
\eea
It turns out that the $\mu$ obtained from \eqref{eq:scal-facts} contains an asymptotic $r^{-2}$ term that can be eliminated using the redundancy \eqref{eq:t-psi-mix}, for the specific value
\begin{align}\label{eq:gamoto}
 \gamma =&\, -1 + \frac{m_-}{4} + \frac12\?h_I\?\left(p^I_+ + p^I_- + \sum_\pA P^I_\pA\right) + \frac{a^2-c^2}{8\?m_-} C^{IJK} h_I \? q_{J}\? q_{K} 
 \CR
 &\, + \frac{m_-}2 C^{IJK} h_I \?\sum_{\pA, \pB}  \? \frac{n^\pA_{J}}{R_\pA-a} \? \frac{n^\pB_{K}}{R_\pB-a} + \frac{1}{2}C^{IJK}h_I q_J \sum_\pA  n_K^\pA  \,.
\end{align}
We henceforth proceed with the solution obtained after \eqref{eq:t-psi-mix} with $\gamma$ as in \eqref{eq:gamoto} has been applied.

In order to simplify the analysis, we take the same approach as in \cite{Bena:2015drs} and fix the asymptotic values of $g_{yy}$ and the dilaton. (Note that there is no loss of generality in doing this, since we keep the radius of the $y$ circle explicitly as $R_y$, and since more general asymptotic values of $e^{2\phi}$ can be restored straightforwardly by rescaling.) This results in the following restrictions on the asymptotic constants $l^0$, $l^I$ and $h_I$:
\begin{gather}\label{eq:AsymPar0}
l^0 = l^I  = \frac{1}{2}\,,
\qquad
h_I = 1 \,,
\end{gather}
while we also find convenient to use \eqref{eq:qI-def} to eliminate the parameters $k_I$ in favour of the $q_I$, as
\begin{equation}
 k_I = 1 + q_I\,,
\end{equation}
where $q_I$ are now a triplet of unrestricted real parameters. Given \eqref{eq:AsymPar0}, the fall-off conditions \eqref{eq:as-beh} are imposed by fixing the parameter $q_0$ that appears in the harmonic part of the function $M$ in \eqref{eq:M-def}, as
\begin{align}\label{eq:AsymPar}
q_0 =&\,{} -1 + \frac14\,m_- + \frac{a^2-c^2}{2\,m_-}\,q_{1}\?q_{2}\?q_3 -\frac12\, q_I \Xi^I 
+ \sum_\pA \left( J^\pA - (R_\pA + a)\,q^0_{\pA} \right)
\cr
&\,
+ \sum_\pA \left( \frac{1}{4}\,C^{IJK}q_Iq_Jn_K^\pA 
- m_-\,\frac{n_I^\pA}{R_\pA - a}\,\left( \frac{p^I_+}{a+c} + \frac{p^I_-}{a-c} \right)  \right)\,,
\end{align}
where we defined the shorthand quantity
\begin{equation}
 \Xi^I \equiv p^I_+ + p^I_- + \frac{a^2-c^2}{4\,m_-}\,C^{IJK}q_{J}q_{K} + \sum_\pA P_\pA^I
 + \sum_\pA C^{IJK} n_J^\pA \left( q_K + \frac{m_-}{R_\pA-a}\,\sum_\pB\frac{n_K^\pB}{R_\pB-a}\right) \,,
\end{equation}
which will be useful in the following.

Once the conditions \eqref{eq:AsymPar0}--\eqref{eq:AsymPar} are imposed, the expressions given in Section \ref{sec:solution} produce an asymptotically $\mathds{R}^{4,1}\times S^1$ solution. However, this solution does not yet possess the asymptotics of a single-center black hole in five spacetime dimensions. The reason is that the asymptotic conditions on the metric leave room for the gauge fields to have a more general behaviour at infinity. In order to restrict to black hole asymptotics, one has to introduce the vectors of five-dimensional electric charges, $Q_I$, and the corresponding constants governing the asymptotic fall-off of the scalars, $E_I$, defined as
\begin{align}
 Q_I = &\, 4\,\frac{a^2-c^2}{m_-} \,q_I + 8\,\sum_\pA n_I^\pA -2\,m_- \Xi^I +2\,C_{IJK} \Xi^J \Xi^K \,,
\\
 E_I = &\, 4\,\frac{a^2-c^2}{m_-} \,q_I + 8\,\sum_\pA n_I^\pA +2\,m_- \Xi^I +2\,C_{IJK} \Xi^J \Xi^K \,.
\end{align}
An asymptotic solution describing a single center five-dimensional black hole must satisfy the conditions
\begin{equation}
 E_1^2 - Q_1^2 = E_2^2 - Q_2^2 = E_3^2 - Q_3^2\; ,\label{AsymptoCY}
\end{equation}
or in other words that all the components of the vector $E_I^2 - Q_I^2$ be equal. This only imposes two conditions on the various parameters.

\subsection{Absence of black holes}
\label{sec:no-bh}

As already mentioned in the Introduction, the solvable system under consideration does not allow for single-center black hole solutions. However, one may consider the possibility of obtaining solutions that contain black holes at the special points of the base space: the centers away from the bolt, and the centers at the poles of the bolt. As part of our general regularity analysis, we now provide a simple analysis ruling out this possibility, therefore restricting the range of interesting solutions within this system to smooth horizonless geometries.

In order to have a black hole horizon at a given special point located at $r_*=0$, the six-dimensional metric \eqref{eq:6D-metr}
must be well-behaved around $r_*=0$. This condition requires that the base metric, $\gamma_{ij}$, be regular, and that the series expansions around $r_*=0$ of the metric functions, $W$, $\mu$ and $H_I$, be: 
\begin{equation}\label{eq:bh-scaling}
W \sim \frac{{\rm w}^2}{r_*^2}\,, \qquad 
H_I \sim \frac{h_I}{r_*^2} \,, \qquad 
\mu\sim \frac{{\rm w} J_{\scriptscriptstyle \rm L} \sin\theta}{r_*^3}\,,
\end{equation}
with strictly positive coefficients ${\rm w},\, h_I$. In addition, one must check the regularity of the would-be horizon, and in particular that it has finite area. The horizon area of a five-dimensional extremal black hole is controlled by the combination
\begin{equation}\label{eq:e4U-bh}
 {\rm e}^{-4\,U} = \frac{H_1 H_2 H_3 - \mu^2}{W} = \frac{S^2 + J_{\scriptscriptstyle \rm L}^2 \sin^2\theta}{r_*^4}\,,
\end{equation}
where $16\pi^2 S>0$ is the horizon area. 

We now analyze in turn the centers away from the bolt, and the centers at the poles of the bolt.\\

\vspace{-3mm}
\noindent
{\bf Centers away from the bolt}\\
We start with the centers away from the bolt, so we set $r_*=\Sigma_\pA$ where \centerA  
 denotes any such center. Near any of these centers, the base metric is smooth by construction, so we need only consider the metric functions. It is a cumbersome but straightforward exercise to expand $W$, $\mu$ and $H_I$ for the solution given in Section \ref{sec:solution} around $\Sigma_\pA=0$, and to investigate whether it is possible to obtain the behaviour \eqref{eq:bh-scaling} by imposing restrictions on the parameters of the solution. 

Considering first the highest poles, and using the notation $\;\det n^\pA \equiv n^\pA_1 n^\pA_2 n^\pA_3$\?, we find the behaviour\footnote{Recall that $\gamma$ is the shift that imposes the correct asymptotics in $\mu$, see below Eq.\;\eq{eq:as-beh}.}
\begin{align}\label{eq:GH-smooth-layer1}
W \,=&\; - 8\,m_-^2 J_\pA (\det n^\pA)\,\frac{R_\pA^2 - c^2}{(R_\pA - a)^3}\,\frac{\cos\theta_\pA}{\Sigma_\pA^5} + {\cal O}(\Sigma_\pA^{-4})\,, 
\CR
\mu \,=&\; (1+ \gamma ) W - 8\,\,m_- J_\pA (\det n^\pA)\,\frac{R_\pA}{|R_\pA|}\left(R_\pA - a \right)\,\frac{R_\pA^2 - c^2}{(R_\pA - a)^3}\,\frac{\cos\theta_\pA}{\Sigma_\pA^5} + {\cal O}(\Sigma_\pA^{-4})\,, 
\CR
H_I \,=&\; 2\,J_\pA\, n^\pA_I \,\frac{R_\pA^2 - c^2}{R_\pA - a}\,\frac{\cos\theta_\pA}{\Sigma_\pA^3} + {\cal O}(\Sigma_\pA^{-2})\,.
\end{align}
One could a priori make several choices in order to cancel these poles. However, any restriction on the bolt background parameters, as $m_-$, $a$ or $c$ would either lead to an extremal limit or degenerate the base, so we restrict to fixing only local parameters at the center. One must have $|R_\pA| > c$ in order for the distance from the bolt to make sense, and assuming that not all the components of $n^\pA_I$ vanish, one is forced to set $J_\pA =0$ in order to make the cubic poles of $H_I$ vanish in \eqref{eq:GH-smooth-layer1}. If all the $n^\pA_I$ are zero, one obtains the quartic pole $W\sim \bigl( \cos \theta_\pA J_\pA ( R_\pA +a)\bigr)^2/ \Sigma_\pA^{\; 4}$, so that indeed one must set $J_\pA =0$.

Continuing with the next-order poles, using the condition $J_\pA =0$ in order to simplify the result, we find the following structure:
\begin{align}\label{eq:GH-smooth-layer2}
W =&\, m_-^2\left(m_-^2 (\det n^\pA)^2\,F_1(R_\pA,m_-,a,c)\,\cos^2\theta_\pA + F_W^\pA \right)\frac{1}{\Sigma_\pA^4} + {\cal O}(\Sigma_\pA^{-3})\,, 
\CR
\mu =&\, m_- \left(m_-^2 (\det n^\pA)^2\,F_2(R_\pA,m_-,a,c)\,\cos^2\theta_\pA + F_\mu^\pA \right)\frac{1}{\Sigma_\pA^4} + {\cal O}(\Sigma_\pA^{-3})\,, 
\CR
H_I =&\, m_- \left(m_- (\det n^\pA)\,n^\pA_I \,F_3(R_\pA,m_-,a,c)\,\cos\theta_\pA + F_I^\pA \right)\frac{\cos\theta_\pA  }{\Sigma_\pA^2} +\frac{\tilde H^\pA_I}{\Sigma_\pA^2} + {\cal O}(\Sigma_\pA^{-1})\,,
\end{align}
where the $F_k(R_\pA,m_-,a,c)$ for $k=1,2,3$ are three independent functions of the quantities displayed in the argument, while the $F_W^\pA$, $F_\mu^\pA$, $F_I^\pA$, $\tilde H^\pA_I$ are independent functions of the same quantities as the $F_k$, the variables $p^I_\pm$ at the bolt and $P^I_\pA$ at the center. The explicit expressions for these functions are rather cumbersome, and are not needed for the present argument; the fact that the $F_k$ are functionally independent means that the only way to remove the unwanted poles proportional to $\cos^2\theta_\pA$ in \eqref{eq:GH-smooth-layer2} using only local variables at center \centerA is to set $n^\pA_1 n^\pA_2 n^\pA_3 = 0$. We therefore impose this, so that the $n^\pA_I$ are rank-2 vectors at each center. For the purpose of exposition, and without loss of generality, we implement this by setting $n^\pA_3=0$. Then \eqref{eq:GH-smooth-layer2} reduces to

\begin{align}\label{eq:GH-smooth-layer25}
W
=&\, m_-^2\,\frac{\left( n^\pA_1 P^1_\pA - n^\pA_2 P^2_\pA \right)^2}{(R_\pA - a)^2\,\Sigma_\pA^4} + {\cal O}(\Sigma_\pA^{-3})\,, 
\CR
\mu  
 =&\, (1+\gamma) W + m_- \frac{R_\pA}{|R_\pA|}\left(R_\pA - a   \right) \frac{\left( n^\pA_1 P^1_\pA - n^\pA_2 P^2_\pA \right)^2}{(R_\pA - a)^2 \, \Sigma_\pA^4}  + {\cal O}(\Sigma_\pA^{-3})\,, 
\CR
H_I  
=&\, 2\,m_- \{ -n^\pA_1 \,,\,\, n^\pA_2 \,, \,\, 0 \}\,
\frac{R_\pA}{|R_\pA|}\frac{\left(n^\pA_1 P^1_\pA - n^\pA_2 P^2_\pA  \right)}{(R_\pA - a)^2}
\frac{\cos\theta_\pA  }{\Sigma_\pA^2} +\frac{H^\pA_I}{\Sigma_\pA^2}  + {\cal O}(\Sigma_\pA^{-1})\,,
\end{align}
where the $H^\pA_I$ are the appropriate restriction of the $\tilde H^\pA_I$ in \eqref{eq:GH-smooth-layer2}. We therefore find that the antisymmetric combination $n^\pA_1 P^1_\pA - n^\pA_2 P^2_\pA$ controls all the unwanted poles and must vanish. The general solution to this equation can be parametrized by (the term proportional to the $q_I$ is added for later convenience)
\begin{equation}\label{eq:p-redef-2}
P^I_{\pA} = C^{IJK}\, n_{J}^\pA (p_K^{\pA}-q_K) \,,
\end{equation}
where the arbitrary component $P^3_\pA$ is parametrized by both $p_1^\pA$ and $p_2^\pA$; this is arranged to ensure that there will be no loss of generality when $n^\pA_I$ is constrained to be rank 1, as it will be shortly. This parametrization is invariant under the further shift $p_1^\pA \rightarrow  p_1^\pA + \epsilon  n_1^\pA  , \, p_2^\pA \rightarrow  p_2^\pA - \epsilon n_2^\pA  $. 
We thus henceforth adopt the redefinition \eqref{eq:p-redef-2}, as anticipated in Eq.\;\eqref{eq:p-redef}. 

At this stage the $H_I$ now have the desired behaviour described in \eqref{eq:bh-scaling}, while both $W$ and $\mu$ still contain $\Sigma_\pA^{-3}$ poles, which we now consider. 
In the interest of brevity we suppress in the following analysis the terms proportional to $J_{\pm}$ and $p^I_{\pm}$, anticipating our later result that $J_{\pm}=p^I_{\pm}=0$ for any regular solution.
The functions $W$ and $\mu$, together with the $\Sigma_\pA^{-2}$ poles of the $H_I$ when the $n^\pA_I$ are rank-2 vectors, then take the form
\begin{align}\label{eq:GH-smooth-layer3}
W =&\,2\,m_-  \frac{R_\pA}{|R_\pA|}\frac{ \left( (R_\pA^{\; 2}-a^2)\,\tilde p^\pA_3 +(a^2 -c^2)\tilde q^\pA_3 \right)}{(R_\pA-a)\,(R_\pA^2-c^2)} \,n^\pA_1\,  n^\pA_2\,\frac{\tilde q^0_{\pA}}{\Sigma_\pA^3} + {\cal O}(\Sigma_\pA^{-2})\,, 
\CR
\mu =&\, \biggl( 1+ \gamma + \frac{R_\pA}{|R_\pA|} \frac{R_\pA-a}{ m_-} \biggr) W -  \frac{n^\pA_1  n^\pA_2 \tilde p^\pA_3 \tilde q^0_{\pA}  }{\Sigma_\pA^3} + {\cal O}(\Sigma_\pA^{-2})\,, 
\CR
H_I =&\, \{ -n^\pA_1 \tilde q^0_{\pA}\,,\,\, - n^\pA_2  \tilde q^0_{\pA}\,, \,\, n^\pA_1\,  n^\pA_2\, (\tilde  p^\pA_3)^2 \}\,\frac{1}{\Sigma_\pA^2}  + {\cal O}(\Sigma_\pA^{-1})\,,
\end{align}
where we used the shorthand definitions
\bea\label{eq:tilde-pars}
\tilde  p^\pA_3 &\equiv&p^\pA_3 - \frac{R_\pA}{|R_\pA|} \frac{ m_-}{R_\pA-a}  \biggl( 1- 2\sum_{\pB\ne \pA} \frac{{\rm sign}(R_\pA-R_\pB) n_{3\pB}}{R_\pB-a} \biggr) \,,  \nn\\
\tilde q^0_{\pA} &\equiv&  2 \frac{ (R_\pA^{\; 2}-c^2)}{R_\pA-a}  q^0_{\pA}  -\frac{1}{2} C^{IJK} ( p^\pA_{I} -q_I)( p^{\pA}_{J} -q_J) n^\pA_K  \nn\\
\tilde q^\pA_3 &\equiv &  q_3 + \frac{R_\pA}{|R_\pA|} m_- \biggl( \frac{R_\pA + a}{a^2-c^2} +2 \sum_{\pB\ne \pA} \Scal{\frac{1}{R_\pB-a} + \frac{R_\pA+a}{a^2-c^2}} \frac{n_{3\pB}}{|R_\pB-R_\pA|}  \biggr) \,.
\eea
Setting either of  $n^\pA_1,\, n^\pA_2,\, \tilde q^0_{\pA} $ to zero would also set to zero the double pole of one component of the $H_I$. Therefore, the only possibility to cancel the cubic pole of $W$ without reducing the rank of the double pole of $H_I$ is to set  $  (R_\pA^{\; 2}-a^2)\,\tilde p^\pA_3 +(a^2 -c^2)\tilde q^\pA_3  =0$. However, from the form of $\mu$ in \eqref{eq:GH-smooth-layer3}, we see that cancelling the cubic pole in $W$ automatically implies that $H_1 H_2 H_3 - \mu^2 \sim {\cal O}(\Sigma_\pA^{-5})$, and therefore that the horizon area vanishes. This implies that \eqref{eq:bh-scaling} and \eqref{eq:e4U-bh} cannot be satisfied. We therefore conclude that it is not possible to obtain a regular extremal black hole with finite horizon located at a finite distance from the non-extremal bolt. \\

\vspace{-3mm}
\noindent
{\bf Poles of the bolt}\\
We now turn to the poles of the bolt, where the metric behaves the same way as the centers away from the bolt, analyzed above.
The analysis is the same for both the North and South pole, so we write $r_*=r_\pm$, and expand for small $r_*$. To do this we introduce coordinates centered on the North / South pole via
\be
r =\, \frac{1}{2} \Scal{ r_\pm + \sqrt{ r_\pm^{\; 2} \pm 4 c \, r_\pm \cos \theta_\pm + 4 c^2}} \,,
\quad
\cos \theta =\, \pm\frac{1}{2c} \Scal{ r_\pm- \sqrt{ r_\pm^{\; 2} \pm 4 c\,  r_\pm \cos \theta_\pm + 4 c^2}} \, . 
\ee
Then near the poles, the three-dimensional base metric $\gamma_{ij}$ behaves as 
\begin{equation}
\gamma_{ij} dx^i dx^j  \sim \varpi_\pm(\theta_\pm)( dr_\pm^{\; 2} + r_\pm^2 d\theta_\pm^{\; 2})+  r_\pm^{\; 2}  \sin^2 \theta_\pm d\varphi^2 \,, 
\end{equation}
with the function 
\begin{equation} \label{eq:varpi}
\varpi_\pm(\theta_\pm) \equiv  \frac{ a^2 + c^2 \mp ( a^2 -c^2 ) \cos\theta_\pm}{2c^2}  \ . 
\end{equation}
Up to this $\theta$-dependent factor, which reduces to unity in the BPS limit, the behaviour of the various functions required for the existence of a black hole horizon is again that in \eqref{eq:bh-scaling}. 

Computing the expansions of $W$ and $\mu$, one obtains 
\begin{eqnarray}\label{eq:pole-smooth-layer1}
W &=& \biggl( \frac{(a\pm c)^2\, ( a \mp c \mp (a\pm c) \cos\theta_\pm)^2 }{4\, c^4\, \varpi_\pm(\theta_\pm)^2}\, J_\pm^{\; 2} \mp\frac{a\pm c}{c\,  \varpi_\pm(\theta_\pm)}\, m_-  p_\pm^1 p_\pm^2 p_\pm^3\biggr) \frac{1}{r_\pm^{\; 4}}  + {\cal O}(r_\pm^{\; -3}) \,,
\CR
\mu &=& \frac{m_- \mp a+c}{m_-} \, W + {\cal O}(r_\pm^{\; -3}) \ , 
\end{eqnarray}
and therefore one must separately impose $J_\pm= 0$ and $ p_\pm^1 p_\pm^2 p_\pm^3 = 0$ for the quartic poles of these functions to vanish. 

We set $J_\pm= 0$ and without loss of generality we choose $p_\pm^3= 0$. 
Having done this, one can examine the cubic poles of $W$ and $\mu$ and the quadratic poles of the $H_I$ at $r_\pm=0$, and find expressions parallel to those in \eqref{eq:GH-smooth-layer3}. The cubic pole of $W$ can be eliminated by either demanding $p_\pm^2 = 0$ (so that the vector $p_\pm^I$ is rank-1), or by solving a linear equation for one of the other parameters (say $p_\mp^I$) associated to the antipodal pole. However, exactly as for the centers away from the bolt, setting the vector $p_\pm^I$ to be rank-1 also eliminates some of the quadratic poles in $H_I$. Similarly, the alternative choice of solving for an antipodal charge $p_\mp^I$ implies that the sextic pole of $H_1 H_2 H_3 - \mu^2$ vanishes, as before. Therefore, we rule out the possibility of a solution with a finite-size black hole  horizon at the poles of the bolt.

\section{Conditions for smooth solutions}
\label{sec:smooth}

We turn now to the analysis of the conditions required for obtaining globally-hyperbolic smooth solutions from our general solution in Section \ref{sec:solution}. The general analysis implies three sets of constraints:  The first set comprises algebraic relations between various parameters, the second set involves a set of inequalities, and the last set involves quantization conditions on specific combinations of the parameters. 

The algebraic equations on the parameters of the solutions follow from the absence of curvature singularities or event horizons at the special points, and the absence of Dirac--Misner string singularities between the special points. The inequalities are the positivity conditions for the dilaton and the signature of the metric to be the same at each special point, and for the absence of closed time-like curves. Finally, the absence of singularities requires the metric to be well-defined on each local patch, with meshing maps that preserve the periodicity of the angular coordinates. This gives quantization conditions on the parameters, as well as arithmetic constraints to avoid orbifold singularities. 
For further details we refer to \cite{Bena:2015drs}, in which this complete analysis was carried out for an explicit example smooth solution containing a non-extremal bolt interacting with one Gibbons--Hawking center (a solution which is of course contained in the present system).
While the first set of algebraic equations can be dealt with systematically, the second and the third sets of constraints can in practice only be analyzed case by case. In this paper we focus on the first set of constraints and leave the analysis of the other constraints (and hence the full construction of new explicit smooth horizonless solutions) for future work. 

In Section \ref{sec:curv-sing} we derive and solve the algebraic constraints associated to the absence of curvature singularities or event horizons at the special points. In Section \ref{sec:CTCs} we derive the algebraic constraints ensuring the absence of Dirac--Misner string singularities between the centers and the vanishing of the Kaluza--Klein vector $\omega$ on the bolt, which is also required for the absence of closed time-like curves. The latter equations define a set of ``bubble equations'' involving the positions of the centers that resembles the corresponding conditions for absence of Dirac--Misner string singularities in analogous extremal solutions. However, we shall see that these ``non-extremal bubble equations'' are much more complicated, and we do not discuss their solution in this paper.

\subsection{Local smooth geometry} \label{sec:curv-sing}

In order to have local smooth geometry at a center,  $r_*=0$, a necessary condition is that the metric functions $W$, $\mu$ and $H_I$ behave as:
%
%
%
\begin{equation}\label{eq:pole-scaling}
W = \frac{W^2_*(\theta)}{r_*^2} + {\cal O}\Scal{\frac{1}{r_*}}\,, \qquad 
H_I = \frac{h_{I*}(\theta)}{r_*} + {\cal O}\big( r_*^0 \big) \,, \qquad 
\mu= {\cal O}\Scal{\frac{1}{r_*}}\,,
\end{equation}
where $W^2_*(\theta)$ and $h_{I*}(\theta)$ are strictly positive functions of $\theta$. 

When the special points are away from the bolt, the function $W^2_*(\theta)$ is a constant, and moreover is the square of an integer (the Gibbons--Hawking charge); we therefore write it as 
\begin{equation}\label{eq:Na-def}
 W = \frac{N_\pA^2}{\Sigma_\pA^2} + {\cal O}(\Sigma_\pA^{-1})\,, \qquad\quad N_\pA \in \mathds{Z} \,.
\end{equation}
For such a center, the local five-dimensional spatial geometry is that of  a $\mathds{Z}_{|N_\pA|}$ quotient of $\mathds{R}^4 \times S^1$. This is a simple generalization of what is known as a Gibbons--Hawking center in four spatial dimensions; for ease of notation we will simply refer to this as a Gibbons--Hawking center. For more details, see the discussion in~\cite{Bena:2015drs}.

At the poles of the bolt, a factor of $\varpi_\pm(\theta_\pm) $ defined in Eq.\;\eq{eq:varpi} again enters, and we have
\begin{equation}\label{eq:Npm-def}
 W = \biggl( \frac{N_{\pm}}{\varpi_\pm(\theta_\pm)\, \Sigma_\pm }\biggr)^2+ {\cal O}(\Sigma_\pm^{-1})\,, \qquad\quad N_\pm \in \mathds{Z} \,.
\end{equation}
%
%

In order to impose \eqref{eq:pole-scaling}, one must first cancel the higher-order poles analyzed in the preceding section. We must therefore impose \begin{equation}\label{eq:pole-cond-1}
 J_\pm = J^\pA = 0 \,,
\end{equation}
as explained below \eqref{eq:GH-smooth-layer1} and  \eqref{eq:pole-smooth-layer1}. Concentrating first on the centers away from the bolt, recall that canceling the higher poles in \eqref{eq:GH-smooth-layer25} moreover requires $n_{I}^{\pA}$ to vanish along one component and $P^I_{\pA}$ to be parametrized as  \eqref{eq:p-redef-2}, leading to the pole structure \eqref{eq:GH-smooth-layer3}. 

The requirement that the quadratic poles of $H_I$ vanish also removes the cubic poles of $W$ and $\mu$, so we focus on the $H_I$. The quadratic poles of $H_I$ can be set to zero in three ways: (i) all three $n_{I\pA}$ vanish; (ii) two  $n_{I\pA}$ vanish and $\tilde q^0_{\pA}$ in \eqref{eq:tilde-pars} vanishes; (iii) both $\tilde q^0_{\pA}$ and $\tilde p_{3\pA}$ in \eqref{eq:tilde-pars} vanish. However, setting $\tilde q^0_{\pA}=\tilde p_{3\pA}=0$ also cancels the first order pole of $H_3$, and respectively for the three other choices, so that option (iii) must be disregarded. In option (i), $n_{I\pA}=0$, one must relax the ansatz \eqref{eq:p-redef}  for $P^I_\pA$ to be non vanishing, and one finds that $P^I_\pA$ must be rank 1 in order for the quadratic poles in $H_I$ to vanish. This solution can nonetheless be considered as a degenerate limit of option (i) in which $n_\pA=0$, so we shall not consider it independently. 

We therefore concentrate on option (ii), which sets the $n_I^\pA$ to be of rank-1 at each center:	
\begin{equation}\label{eq:n-rank-1}
 C^{IJK} n_J^\pA n_K^\pA = 0 \,,
\end{equation}
and to impose 
\begin{equation}\label{eq:q0A-sol}
 q^0_{\pA} = \frac14\,\frac{R_\pA -a}{R_\pA^2 - c^2} \,C^{IJK} n_I^\pA (p_J^\pA-q_J)(p_K^\pA-q_K) \, . 
\end{equation}
Considering the definition of the $P_\pA^I$ in \eqref{eq:p-redef} along with the condition \eqref{eq:n-rank-1}, we note that the component of the $p_I^\pA$ along the direction of $n^I_\pA$ does not appear in the solution. The pole at the center 
\centerA
is therefore eventually parametrized by one non-zero component of $n^I_\pA$ and two components of $p_I^\pA$. This is also in agreement with the solutions found in  \cite{Bena:2015drs}, in which the ansatz assumed that only $n_3^\pA\ne 0$.\footnote{Since that solution was given in the context of a different parametrization for the system, a complete translation to the language of this paper is a cumbersome but straightforward task; the map is given in App.~\ref{sec:map-to-old}.}

In the BPS limit, this solution reproduces the behaviour of a standard smooth supersymmetric Gibbons--Hawking center, with 
\be{\cal H}^0 \sim \frac{q^0_{\pA}}{\Sigma_\pA} \ ,  \qquad {\cal H}^I \sim \frac{P^I_{\pA}}{\Sigma_{\pA}}   \ , \qquad {\cal H}_I \sim \frac{n_{I\pA}}{\Sigma_{\pA}} \ , \quad  {\cal H}_0 \sim 0 \ , \label{BPSlimCharge} \ee
up to overall normalization factors. Note that this is the S-dual of a supertube center (see for example~\cite{Mateos:2001qs,Bena:2008dw}). 

We note that the integer Gibbons--Hawking charge appearing in Eq.\;\ref{eq:Na-def} is given by
\begin{align}\label{eq:Wa}
N_\pA = 
 &\, \frac{1}{2}\,\sum_{I} n_I^\pA p_{I+1}^\pA p_{I+2}^\pA 
      - \frac{1}{2}\,\frac{a^2 - c^2}{R_\pA^2 - c^2} \,\sum_{I} n_I^\pA (p_{I+1}^\pA-q_{I+1})(p_{I+2}^\pA-q_{I+2})
 \CR
  &\, + m_-\sum_{I,\,\pB\neq\pA} \frac{R_\pA}{|R_\pA|}\, \frac{n_{I+1}^\pA n_{I+2}^\pB}{|R_\pA-R_\pB|}\,
           \left( \frac{p_{I}^\pA}{R_\pB-a} - \frac{p_{I}^\pB}{R_\pA-a} \right) .
\end{align}

The same analysis applies in the vicinity of the poles of the bolt. One finds that in order to cancel the double poles of all the $H_I$, the  $p^I_\pm$ must be at most rank 1. A further condition must be implemented, which can be obtained for either $p^I_\pm=0$ or constraining the $q_I$, but the second leads to a cancellation of the first order pole of one of the $H_I$ function. The only consistent solution is therefore to set 
\begin{equation} \label{eq:p-pm-0}
 p^I_\pm = 0\, . 
\end{equation}
Note that this choice implies that in the solution of Section \ref{sec:solution}, there are no remaining parameters that are intrinsic to the poles of the bolt (the parameters $a,c,m_-,q_I$ are associated to the bolt as a whole, rather than to its poles). One determines then the Gibbons--Hawking charges $N_\pm$ at the poles as 
\be N_\pm  = - \frac{a\pm c }{2\, c}\biggl(  1+ \sum_\pA N_\pA \biggr) \mp \frac{x}{2} \ , \label{GHpoles}  \ee
where we introduced the integer
\begin{align} \label{eq:x-def}
 x = &\,\frac{a^2 - c^2}{2\,m_-}\,q_1q_2q_3 
 \CR
 &\, + \frac{a^2 - c^2}{2\,c}\,\sum_{I,\,\pA} \frac{R_\pA}{|R_\pA|} \frac{n_I^\pA}{R_\pA -a} \,\left( \frac{(R_\pA -a)^2}{R_\pA^2 - c^2}\,(p_{I+1}^\pA-q_{I+1})(p_{I+2}^\pA-q_{I+2}) -q_{I+1} q_{I+2} \right)\, . 
\end{align}
Note that the Gibbons--Hawking charge is only additive in the extremal limit in which $a=c$, such that $N_++N_- + \sum_\pA N_\pA = -1$, because the bolt is only regular in six dimensions for an integer $\frac{a}{c}$ greater than $1$. The two  Gibbons--Hawking charges must be integral, implying that $x$ must be an integer with the same parity as $\frac{a+c}{c}( 1+\sum_\pA N_\pA)$.

\subsection{Absence of closed time-like curves}\label{sec:CTCs}

We finally examine the constraints arising from the absence of closed time-like curves.  While this is hard to do in general since it requires a careful analysis of the global properties of solutions, a first strong requirement is that the vector $\omega$, describing the time fibration, is globally defined over the space-like base. Given that the solution under consideration is axisymmetric, the global definition of $\omega$ amounts to the condition that $\omega_\varphi$ is continuous on the symmetry axis. 
 
It is straightforward to use the expression in Appendix \ref{sec:vec-fields} together with the restrictions \eqref{eq:pole-cond-1}--\eqref{eq:p-pm-0} to compute the potential discontinuities of $\omega$ at the bolt and at the Gibbons--Hawking centers. Note that Eq.\;\eqref{eq:AsymPar} already ensures that $\omega$ is single-valued at asymptotic infinity, so we only need to impose its continuity at the special points in the bulk. This leaves us with potential discontinuities at the Gibbons--Hawking centers and a potential discontinuity on the bolt. 

%

In order to write the conditions for the the vector field $\omega$ to be continuous, let us introduce the following shorthand quantity, that will also be useful below:
\begin{equation}\label{eq:C0-def}
C_{0} \equiv  \frac{m_-}{2} + \frac{a^2 - c^2}{2\,m_-}\,\sum_{I} q_{I+1} q_{I+2}
 + \sum_{\pA, I \ne J} n_I^\pA p_{J}^\pA + 2\,m_-\!\!\!\sum_{I,\,\pA,\,\pB} \frac{n_{I+1}^\pA}{R_\pA -a}\frac{n_{I+2}^\pB}{R_\pB -a} \,,
\end{equation}
as well as the sign, $\varepsilon_{\pA \pB}$, depending on the position of centers
\begin{equation}
 \varepsilon_{\pA \pB} \equiv \frac{R_\pA-R_\pB}{|R_\pA-R_\pB|} \,.
\end{equation}
In terms of these quantities and the Gibbons--Hawking charges $N_\pA$ given in \eqref{eq:Wa}, the conditions required for the discontinuities of the vector field $\omega$ to vanish at the Gibbons--Hawking centers are given by\footnote{Note that we use the rank 1 condition \eqref{eq:n-rank-1} of $n^I_\pA$ to simplify these formulae.}
\begin{align}\label{eq:jumps}
C_{0} \, N_\pA +&\,  \sum_{I\ne J} n_I^\pA  p_{J}^\pA  +2 \sum_{I,\,\pB\neq\pA} n_{I+1}^\pA n_{I+2}^\pB \frac{p_{I}^\pA-p_{I}^\pB}{|R_\pA-R_\pB|}  
 \CR
 =&\,   \ \frac{R_\pA}{|R_\pA|} \frac{a^2 - c^2}{m_-}\frac{R_\pA -a}{R_\pA^2 - c^2} \,\sum_{I} n_I^\pA (p_{I+1}^\pA-q_{I+1})(p_{I+2}^\pA-q_{I+2})
 \CR
  &\, +m_- \frac{R_\pA}{|R_\pA|}\sum_{I} \frac{n_I^\pA}{R_\pA -a}\,\left(1 -2\sum_{\pB} \varepsilon_{\pA \pB}\,\frac{n_{I+1}^\pB}{R_\pB -a} \right)\,\left(1 -2 \sum_{\pC}\varepsilon_{\pA \pC}\,\frac{n_{I+2}^\pC}{R_\pC -a} \right)\,.
\end{align}

We now consider the continuity of $\omega$ at the bolt. 
The coordinate $\varphi$ degenerates only on the poles of the bolt, not everywhere on the bolt;
however $\omega_\varphi$ is constant on the bolt, and so must vanish identically on the bolt by continuity \cite{Bena:2015drs}. 
One can also verify that the same condition $\omega_\varphi|_{B}=0$ is required for the quadratic pole of the function $\mu$ to vanish on the poles of the bolt. 
Simplifying the expression of $\omega_\varphi|_{B}$ assuming  that \eqref{eq:AsymPar} and \eqref{eq:jumps} hold, one obtains:
\begin{eqnarray}
 \frac{a}{c}\,\omega_\varphi\big|_{B} 
&=&\, \frac{a^2 - c^2}{c\,m_-}\,\left( \sum_\pA N_\pA +1 + \frac{a^2 - c^2}{4\,m_-}\,q_1q_2q_3   \right)- C_0\, \left( \frac{x}{2} + \sum_\pA N_\pA +1 \right)
 \CR
&& {} + \frac{a^2 - c^2}{4\, c} \sum_{I}  q_I    \left(1 +2\sum_{\pA}\frac{R_\pA}{|R_\pA|} \frac{n_{I+1}^\pA}{R_\pA -a} \right)\,\left(1 +2 \sum_{\pB} \frac{R_\pB}{|R_\pB|} \frac{n_{I+2}^\pB}{R_\pB -a} \right)  \label{eq:omega-cons} \\
&& {} +   \frac{m_-}{2}\left(1 +2 \sum_{\pA}\frac{R_\pA}{|R_\pA|}\frac{n_1^\pA}{R_\pA -a} \right)\left(1 +2\sum_{\pB} \frac{R_\pB}{|R_\pB|}\frac{n_{2}^\pB}{R_\pB -a} \right)\,\left(1 +2 \sum_{\pC}\frac{R_\pC}{|R_\pC|} \frac{n_{3}^\pC}{R_\pC -a} \right) , 
\nonumber
\end{eqnarray}
where  $x$ is the integer defined in \eqref{eq:x-def}. 

The vanishing of the expressions \eqref{eq:jumps} and \eqref{eq:omega-cons} is the analogue of the bubble equations appearing in extremal solutions, both BPS and non-BPS alike \cite{Denef:2000nb, Bena:2009en,Bossard:2013nwa}. Indeed, imposing the BPS limit $a\rightarrow c,m_-\rightarrow 0$ with $\frac{a^2 -c^2}{m_-}$ kept fixed as in Section \ref{sec:extr-limits}, one finds that these constraints reduce to the BPS bubble equations for a set of Gibbons--Hawking centers defined by the harmonic functions $\cH^\Lambda$, $\cH_\Lambda$ of \eqref{eq:BPS-har} with restricted poles  according to \eqref{BPSlimCharge}. It is in particular straightforward to see that these equations become linear in the inverse distances. This is consistent with the fact that the bolt reduces to a pair of Gibbons--Hawking centers in the BPS limit \cite{Giusto:2004id,Giusto:2004ip,Giusto:2012yz}. The connection to the Almost--BPS system in the limit given in Section \ref{sec:extr-limits} is less straightforward, as it leads to a non-standard duality frame. In the almost-BPS extremal limit, the poles of the ansatz functions at the South pole of the bolt turn out to vanish identically (in particular $N_-=0$ in \eqref{GHpoles}) .

It is also important to compute the value of the vector $w^0$ defining the fibration over $\psi$ on the bolt,
\begin{align}\label{eq:w0-bolt}
 \frac{a}{c}\,w^0_\varphi \big|_{B} =&\, 1 + x + \sum_\pA \left( 1 - \frac{a}{c}\, \frac{R_\pA}{|R_\pA|} \right) N_\pA \ ,
\end{align}
since the regularity conditions at the bolt imply  that 
\be \frac{a}{c} = m-n \ , \qquad \frac{a}{c}\,w^0_\varphi \big|_{B} = m+n \ , \ee
for two integers $m$ and $n$~\cite{Jejjala:2005yu,Bena:2015drs}. One then finds that the Gibbons--Hawking charges at the poles of the bolt are automatically integers:
%
\be N_+ \,=\, -m - \sum_{R_\pA>c}  N_\pA \ , \qquad N_- \,=\, n -   \sum_{R_\pA< -c}  N_\pA \ .  \ee

Let us summarize the set of free parameters in our solutions and the physical/geometrical quantities they correspond to. In the following we switch back to discussing solutions of general six-dimensional supergravity theories with $\nt$ tensor multiplets. One can consider that $c$ and $R_\pA$ are determined by the bubble equations \eqref{eq:jumps} and \eqref{eq:omega-cons}. Then the parameters  $q_I,\, a,\, m_-$ at the bolt are understood to parametrize the two integers $m$ and $n$ characterizing its topology, the flux $Q_a = \frac{1}{4\pi^2} \int_B G_a$ over the bolt 3-cycle, and the radius $R_y$ associated to the $y$ coordinate. Each new Gibbons--Hawking center is a $\mathds{Z}_{|N_\pA|}$ quotient of $\mathds{R}^4 \times S^1$ parametrized by two integers $N_\pA$ and $M_\pA$ (for the action on the additional circle),  and its presence introduces one new 3-cycle that supports $\nt+1$ fluxes $F^\pA_a = \frac{1}{4\pi^2 }\int_{\Sigma_\pA}G_a$ \cite{Bena:2015drs}.
We thus see that each Gibbons--Hawking center is parametrized by $\nt+2$ parameters $n^\pA_I, \, p^\pA_I$ for $\nt+3$ new physical quantities $N_\pA,\, M_\pA,\, F^\pA_a$. We therefore understand that the additional integers $M_\pA$ (say) can be thought of as determined in terms of the other quantities. 
Given integer values for $m,\, n,\, Q_a,\, N_\pA$ and $F^\pA_a$, it would be nice if the $M_\pA$ were automatically integers, however this is rather difficult to check. 
Moreover, for the solution to have the same asymptotics as a Cvetic--Youm black hole, one must constrain $\nt+1$ additional parameters to satisfy \eqref{AsymptoCY}. 
Therefore, on one of the Gibbons--Hawking centers, the fluxes $F^\pA_a$ must be determined in terms of other parameters. 
For a single additional center, the only free parameter is its Gibbons--Hawking charge $N_1$, as in the solution derived in  \cite{Bena:2015drs}. 

It will be very interesting to explore the space of non-extremal smooth horizonless supergravity solutions contained in our general solution. Work in this direction is in progress.

\section*{Acknowledgements}

We thank Amitabh Virmani and Nick Warner for valuable discussions. SK is grateful to CPHT, Ecole Polytechnique, for hospitality in the early stages of this work. This work was supported in part by the Agence Nationale de la Recherche grant Black-dS-String. The work of IB, SK and DT was supported in part by the John Templeton Foundation Grant 48222. The work of SK was supported in part by INFN and by the European Research Council under the European Union's Seventh Framework Program (FP/2007-2013) -- ERC Grant Agreement 307286 (XD-STRING). The work of DT was supported by a CEA Enhanced Eurotalents Fellowship.

\appendix

\section{Relation to the Floating JMaRT system}
\label{sec:map-to-old}

In this appendix we briefly describe the relation of the solvable system given in Section \ref{sec:system} to the ``Floating JMaRT'' system constructed in \cite{Bossard:2014ola} and used in \cite{Bena:2015drs} to obtain an explicit solution with a single Gibbons--Hawking center together with a smooth bolt. The system in \cite{Bossard:2014ola} is based on a set of Ernst potentials for an auxiliary Euclidean Maxwell-Einstein solution, denoted by $\cE_\pm$ and $\Ppm$, together with six more functions, $L^I$ and $K_I$, for $I=1,2,3$, where there was no explicit triality covariance despite the naming. To avoid confusion with the functions of the same name appearing in this paper, we will use the notation $L^I{}^\old$ and $K_I{}^\old$ for the functions appearing in \cite{Bossard:2014ola}.

Explicitly, in terms of the functions appearing in the ansatz in Section \ref{sec:system}, we have the identifications
\begin{gather}
 \cE_+  =  -\frac{K_3}{2 + K_3 + 2\, \Vb}\,, \qquad \cE_-  =  \frac2{V} - 1 \,, 
\CR
 \Pp  =  \lambda \,\frac{1 + K_3 + 2\, \Vb }{2 + K_3 + 2\, \Vb} \,, \qquad \Pm  =  \frac{2}{\lambda}\,\frac{V - 1}{V}\,,
 \label{eq:Ernst-pot}
\end{gather}
for the Ernst potentials, where $\lambda$ is a free parameter, set to $\lambda=m_-/e_-$ when comparing with the explicit Maxwell-Einstein solution used in \cite{Bossard:2014ola}. Note that this is not an honest redefinition, since the four original Ernst potentials are mapped to only three functions. This is a particular choice inspired by the fact that the nontrivial Maxwell-Einstein solutions we use are such that $\cE_-$ and $\Pm$ are those of an extremal solution and are therefore not independent.

We then proceed to the $L^a{}^\old$ and $K_a{}^\old$, for $a=1,2$ and using the $\eta_{ab}$ in \eqref{Etaab}, for which we find
\begin{equation}
 L^a{}^\old  =  \frac12\,\frac{\eta^{ab} K_b + L^a}{2 + K_3 + 2\, \Vb}\,, 
 \qquad 
 K_a{}^\old  =  \frac{1}{2\,\lambda}\,\left( K_a - 2\,(\Vb+1)\,\frac{K_a + \eta_{ab} L^b}{2 + K_3 + 2\, \Vb}  \right)\, . 
\end{equation}
The final two functions in the Floating JMaRT system are identified as 
\begin{align}
 K_3{}^\old  =  &  \frac{1}{4\,\lambda}\,\left( M + L^3 - \frac12\,\frac{V}{1+V\Vb}\,\left( K_a L^a + K_1 K_2 \right)
 +\frac{V-1}{1+V\Vb}\,\frac{(K_1 + L^2)\,(K_2 + L^1)}{2 + K_3 + 2\, \Vb}  \right) \,, 
\CR
 L^3{}^\old  =  & - \frac{1}{\lambda^2}\,\left( \frac{\Vb+1}{2}\, M - \frac14\, K_I L^I + \frac18\,\frac{V}{1+V\Vb}\,K_1 K_2 K_3
                                                                                             - \frac12\,\frac{V-1}{1+V\Vb}\,\left( K_a L^a + K_1 K_2 \right) \right.
 \CR
                                           &\qquad\qquad  + \left. \frac{(V-1)\,(\Vb+1)}{1+V\Vb}\,\frac{(K_1 + L^2)\,(K_2 + L^1)}{2 + K_3 + 2\, \Vb}  \right) \,, 
\end{align}
where we caution that we use both a sum over indices $a=1,2$ and $I=1,2,3$ for convenience.

Besides these identifications, we further applied a gauge transformation on the gauge field $A^3 \rightarrow A^3 - 2\, dt + 2 \,\lambda\, d\psi$ and we rescaled all fields appropriately in order to remove the explicit dependence on the parameter $\lambda$. The latter is enforced by a rescaling of the coordinates
\begin{equation}
 t \rightarrow 16\,\lambda\,t\,, \qquad   y \rightarrow 8\,\lambda\,y\,, \qquad    \psi \rightarrow 8\,\lambda^2\,\psi\,,
\end{equation}
while also imposing that the six-dimensional metric and the two-form potentials $C_a$ rescale by a factor of $32\,\lambda^2$, and the dilaton is invariant. We display the redefinitions used explicitly in this paper:
\begin{gather}
 W \rightarrow 64\,\lambda^4\,W\,, \qquad   \mu \rightarrow 128\,\lambda^3\,\mu\,, \qquad  
 \{ H_1,\,H_2,\,H_3 \} \rightarrow 16\,\lambda^2\,\{ 2\;\!H_1,\,2\;\!H_2,\, H_3 \}\,,
 \CR
 \omega \rightarrow 16\,\lambda\,\omega\,, \qquad  \{ w^0,\, w^1,\,w^2,\,w^3 \} \rightarrow 4\,\lambda^2\,\{ 2\;\!w^0,\, w^1,\,w^2,\,2\;\!w^3 \} \,,
\end{gather}
while the rest are fixed uniquely by imposing consistency.
Finally, we flipped the overall sign of the two gauge fields $A^1$ and $A^2$.

\section{Vector fields}
\label{sec:vec-fields}

In this appendix we list the explicit expressions for the various vector fields used in the main text. Since we deal exclusively with axisymmetric solutions, all vector fields only have a single component, along $\varphi$ in the 3D base \eqref{eq:3D-base}, which is displayed explicitly below. We first define some useful functions
\begin{align}
 S(r,\theta) \equiv &\, \frac{\sin^2\theta}{r^2 - c^2 + a^2 \sin^2\theta}\,,
 \CR
 \cW_0 \equiv &\, \cos\theta + a\,S(r,\theta)\,(r-a\,\cos\theta + m_-)\,,
 \CR
 \cW_\pm \equiv &\, \cos\theta \mp a\,S(r,\theta)\,(r \pm a\,\cos\theta)\,,
 \CR
 \cV_\pm \equiv &\, \frac{r\,\cos\theta \mp c}{r \mp c\,\cos\theta}\,,
\end{align}
which we use for brevity. Additionally, we use the shorthand $c_\theta\equiv\cos\theta$ for the remainder of this appendix.

Starting from the electric vector fields, using the general solution \eqref{eq:KI-def}--\eqref{eq:M-def} we find from \eqref{baEq-2}  the $\varphi$-components
\begin{align}
 (v_I)_\varphi = &\, \frac{a^2 - c^2}{m_-}\, q_I\, \cW_0 + h_I \left( \frac{a^2 - c^2}{m_-}\, (\cW_- - \cW_0) - m_- \cW_+ \right)
 \CR
 &\, + 2\,\sum_\pA\frac{n_I^\pA}{(R_\pA-a)\,\Sigma_\pA}\left[ c^2 -R_\pA^2 + m_- (r - R_\pA c_\theta) 
  \right. 
 \CR
 &\, \qquad \qquad \qquad \qquad \qquad \left. 
  + \left((a^2 - c^2)\,c_\theta + (r + a\,c_\theta )\,(R_\pA-a)\right)\,\cW_0\right] \,.
\end{align}
Similarly, from \eqref{wIEq-1} we find the following expression for the $\varphi$-components of the magnetic vector fields $w^I$, $w^0$
\begin{align}
 (w^I)_\varphi = &\, -\frac12\,C^{IJK} q_J\, v_K + \frac{a^2 - c^2}{2\,m_-}\,C^{IJK} q_J\, q_K \,(\cW_0 - \tfrac12\, \cW_-) 
 \CR
 &\, - \frac14\,m_- C^{IJK} h_J(h_K + 2\,q_K)\, \cW_+  - p^I_+\,\cV_+- p^I_-\,\cV_-
 \CR
 &\,     + m_-\sum_\pA \frac{C^{IJK}(q_J+h_J)\,n_K^\pA}{(R_\pA-a)\,\Sigma_\pA}\,
     \left( r - a\,c_\theta - (R_\pA - a)\, \cW_+ + \frac{a^2 - c^2}{2\,a} ( 2\,c_\theta - \cW_+ - \cW_- ) \right)
 \CR
 &\,     + \sum_\pA C^{IJK}n_J^\pA\,(p_K^\pA - q_K)\,\frac{R_\pA-r\,c_\theta}{\Sigma_\pA}
 \CR
 &\,     +m_- \sum_{\pA,\,\pB} \frac{n_{I+1}^\pA n_{I+2}^\pB}{(R_\pA-a)\,(R_\pB-a)} \frac{1}{\Sigma_\pA \Sigma_\pB}\left[ 
        \left( (R_\pA - R_\pB)^2 - \Sigma_\pA^2 - \Sigma_\pB^2 \right)\, \cW_+
        \right.
 \\
 &\, \qquad \qquad \qquad
 \left. + S(r,\theta)\,(r^2 - c^2)\,\left(2\, (r + a\, c_\theta )\,(R_\pA + R_\pB - 2\, a) + 4\, (a^2 - c^2) c_\theta \right) \right]
 \,,\nonumber
\end{align}
\small
\begin{align}
 (w^0)_\varphi =  &\, -\frac18\,C^{IJK}q_I\,q_J\,v_K -\frac12\,q_I w^I + q_0\,\cW_+ 
       + \frac{a^2 - c^2}{m_-}\,\left( l^0  -\tfrac12\, (h_I+q_I) l^I \right)\,(\cW_0-\cW_-)
 \CR
 &\, + \frac{a^2 - c^2}{4\,m_-}\,\left( \left( k_1 k_2 k_3 + 2\, q_1 q_2 q_3 \right)\,\cW_0 
         - \left( k_1 k_2 k_3 + q_1 q_2 q_3 \right)\,\cW_- \right)
     + \frac{m_-}{4}\,\left( q_1 q_2 q_3 - k_1 k_2 k_3 \right)\,\cW_+
 \CR
 &\, - \sum_\pA \frac{q^0_{\pA}}{\Sigma_\pA}\,\left[ 
     \left(R_\pA + a -a\,(a^2 - c^2)\,S(r,\theta) \right)\,(R_\pA - r\,c_\theta )
     +(a^2 - c^2)\,(r^2 - c^2)\,S(r,\theta) \right] 
 \CR
 &\, + \sum_\pA \frac{J_\pA}{\Sigma_\pA}\,\left[
     R_\pA -r\,c_\theta  + \frac{(r^2 - c^2)\,(r + a\,c_\theta)}{\Sigma_\pA^2}\, S(r,\theta) \,
     \left((r - a\,c_\theta) (R_\pA + a) +(a^2 - c^2)\,c_\theta \right)\right]
 \CR
 &\, + \sum_{\epsilon=\pm 1} J_\epsilon
 \left[ \frac1a\,(\cW_+ - \cW_-)-(a+\epsilon\,c)\,\frac{S(r,\theta)}{r - \epsilon\,c \,c_\theta} 
     \left(\left(2\,r+(a-\epsilon\,c)\,c_\theta\right)\cV_\epsilon + r \,c_\theta +  a\right)  \right]
 \CR
 &\,
  - m_-\sum_{\pA,\epsilon=\pm 1}\frac{n_I^\pA\,p^I_\epsilon}{(R_\pA-a)\,\Sigma_\pA}\,
 \left( \frac{R_\pA - r\,c_\theta}{a +\epsilon c} + 2\,S(r,\theta)\,(r^2 - c^2)\,\frac{r + a\,c_\theta}{r -\epsilon\, c\,c_\theta} \right)
 \CR
 &\,
  -m_- \left( l^I - \frac14\,C^{IJK}k_Jk_K \right)\sum_{\pA}\frac{n_I^\pA}{(R_\pA-a)\,\Sigma_\pA}
 \left( r - a\,c_\theta -(R_\pA-a)\, \cW_+ + a\,(a^2 - c^2)\,S(r,\theta)\,c_\theta \right)
 \CR
 &\,
  - m_-\sum_{\pA,\pB}\frac{C^{IJK} n_I^\pA n_J^\pB\,(p_K^\pB - q_K)}{(R_\pA-a)\,\Sigma_\pA\,\Sigma_\pB}\,
  \left[ S(r,\theta)\,(r^2 - c^2)\,\left( r + a\,c_\theta \right) 
      \right. 
 \CR
 &\,  \qquad \quad \quad \qquad \quad \quad \qquad \quad \quad \qquad \quad \quad \quad   
  \left. - \frac12\,\cW_+ 
           \left( R_\pA - R_\pB - \frac{(\Sigma_\pA - \Sigma_\pB)^2}{R_\pA - R_\pB} \right) \right]
 \CR
 &\,
  -4\,m_-^2 \sum_{\pA,\pB,\pC}\frac{n_1^\pA n_2^\pB\,n_3^\pC}{(R_\pA-a)\,(R_\pB-a)\,(R_\pC-a)}\,
   S(r,\theta)\,\frac{(r^2 - c^2)\,(r^2 - c^2c_\theta^2)}{\Sigma_\pA\,\Sigma_\pB\,\Sigma_\pC}
 \,,
\end{align}
\normalsize

Finally, the vector field $\omega$ is also determined from \eqref{wIEq-1} as
\small
\begin{align}
 \omega_\varphi = &\, \frac12\, (h_I + q_I) w^I + \frac18\, C^{IJK}(q_I\,q_J -h_I\, h_J)\,v_K + \frac12\, l^I v_I 
 \CR
 &\, - l^0 \cW_0 -\tfrac12\, h_I l^I\,\cW_+ + \frac{q_0}{m_-}\,(\cW_0 + \cW_+) 
 \CR
 &\,  - \frac12\,(q^1+h^1)\,(q^2+h^2)\,(q^3+h^3)\,\left( \cW_- + \tfrac32\,\cW_0 \right)
     + \frac12\,C^{IJK}h_I q_J(h_K + q_K)\cW_+
 \CR
 &\, + \sum_\pA \frac{q^0_{\pA}}{\Sigma_\pA}\,\left[
     \frac{a^2 - c^2}{m_-}\,\left( r + a\,c_\theta 
     + m_- (1 - c_\theta \cW_+) +\frac{a^2 - c^2}{m_-}\,c_\theta\,(\cW_- - \cW_0)\right) \right.
 \CR
 &\, \qquad \qquad \qquad
  \left. +(R_\pA + a) \left(R_\pA - r\,c_\theta-\frac{a^2 - c^2}{m_-}\,\cW_0\right) \right] 
 \CR
 &\, - \sum_\pA \frac{J_\pA}{\Sigma_\pA}\,\left[
     R_\pA -r\,c_\theta  + \frac{(r^2 - c^2)\,\Sigma_+}{\Sigma_\pA^2}\, S(r,\theta) \,
     \left((r - a\,c_\theta) (R_\pA + a) +(a^2 - c^2)\,c_\theta \right)\right]
 \CR
 &\, + \frac1{a\,m_-}\,(J_+ + J_-)\left(  \frac{a^2 - c^2}{m_-}\,(\cW_- - \cW_0) - m_-(\cW_+ - \cW_-) \right)
 \CR
 &\, + \sum_{\epsilon=\pm 1} J_\epsilon(a+\epsilon\,c)\,\frac{S(r,\theta)}{r-\epsilon\,c\, c_\theta}
 \left[ (a -\epsilon\,c) \,c_\theta \, \cV_\epsilon- r \,c_\theta +  a  
       + \left(2\,r - \frac{a^2 - c^2}{m_-}\right)\, ( c_\theta + \cV_\epsilon ) \right]
 \CR
 &\,
  + \sum_{\pA,\epsilon=\pm 1}\frac{n_I^\pA}{(R_\pA-a)\,\Sigma_\pA}
  \frac{p^I_\epsilon}{r -\epsilon\, c\,c_\theta}\,\left[  c^2 \sin^2\theta -r (R_\pA c_\theta - r) 
    \right. 
 \CR
 &\,  \qquad \qquad \qquad  \qquad \qquad \qquad \qquad  
  \left. + 2\,S(r,\theta)\left( m_-(r^2 - c^2)\,(r +a\,c_\theta) - a^2 (r^2 - c^2c_\theta^2) \right) \right]
 \CR
 &\,
  + \sum_{\pA,\epsilon=\pm 1}\frac{n_I^\pA\,p^I_\epsilon}{\Sigma_\pA}\,\frac{R_\pA - r\,c_\theta}{R_\pA-a}\,
 \left( \frac{m_-}{a +\epsilon c}\, + \epsilon\,\frac{c}{r -\epsilon\, c\,c_\theta} \right)
 \CR
 &\,
  +\frac14\,m_- \sum_{\pA}\frac{C^{IJK}k_I k_J n_K^\pA}{(R_\pA-a)\,\Sigma_\pA}
 \left( r - a\,c_\theta -(R_\pA-a)\, \cW_+ + a\,(a^2 - c^2)\,S(r,\theta)\,c_\theta \right)
 \CR
 &\,
  + \sum_{\pA,\pB}\frac{C^{IJK} n_I^\pA n_J^\pB\,(p_K^\pB - q_K)}{2\,(R_\pA-a)\,\Sigma_\pA\,\Sigma_\pB}\,
  \left[ 2\,m_-S(r,\theta)\,(r^2 - c^2)\,\Sigma_+
      \right. 
 \CR
 &\,  \qquad \quad \quad  
  \left. - \left(m_- \cW_+ - \frac{a^2 - c^2}{m_-}\, (\cW_- - \cW_0) + R_\pA-a\right)\,
           \left( R_\pA - R_\pB - \frac{(\Sigma_\pA - \Sigma_\pB)^2}{R_\pA - R_\pB} \right) \right]
 \CR
 &\,
  +2\,m_- \sum_{\pA,\pB,\pC}\frac{n_1^\pA n_2^\pB\,n_3^\pC}{(R_\pA-a)\,(R_\pB-a)\,(R_\pC-a)}\,
  \left( \cW_{\pA\pB\pC} -2\,m_-S(r,\theta)\,\frac{(r^2 - c^2)\,(r^2 - c^2c_\theta^2)}{\Sigma_\pA\,\Sigma_\pB\,\Sigma_\pC} \right) 
 \,,
\end{align}
\normalsize
where $\cW_{\pA\pB\pC}$ is given by
\small
\begin{align}
\cW_{\pA\pB\pC} \equiv &\, \frac{1}{\Sigma_\pA\,\Sigma_\pB\,\Sigma_\pC} \left[ R_\pA R_\pB R_\pC\,\cW_+ 
          + c^2 \sin^2\theta\,c_\theta\,(R_\pA + R_\pB + R_\pC - a )- r\, (r^2 - c^2)
      \right. 
 \CR
 &\,  \qquad \quad \quad  
      + (a^2 - c^2)\,\sin^2\theta\,\left(r -a\, S(r,\theta)\, (a\, r - c^2\,c_\theta) \right)
 \CR
 &\,  \qquad \quad \quad  
      - \left(r -a\, S(r,\theta)\, (a\, r + c^2\,c_\theta) \right)\times
 \CR
 &\,  \qquad \quad \quad \quad \quad  
  \left. 
      \left( R_\pA R_\pB + R_\pA R_\pC + R_\pB R_\pC - (a\, \sin^2\theta +r\,c_\theta)\, (R_\pA + R_\pB + R_\pC) + c^2\right)
  \right] \,.
\end{align}
\normalsize

\newpage

\bibliographystyle{JHEP} 
\bibliography{PaperG} 

\end{document}